\documentclass[11pt, a4paper]{article}
\pdfoutput=1

\usepackage{amsmath}
\usepackage{amsfonts}
\usepackage{amssymb}
\usepackage{graphicx}
\usepackage{mathrsfs}
\usepackage{latexsym}
\usepackage{color}
\usepackage{cite}
\usepackage{slashed, cancel}
\usepackage{hyperref}
\usepackage{bbold}
\usepackage{soul} 

\usepackage[font=footnotesize,labelfont=bf]{caption} 

\numberwithin{equation}{section}

\definecolor{rossos}{rgb}{0.8,0.2,0.3}
\definecolor{bluscuro}{rgb}{0.15, 0.2, .85}
\definecolor{bluchiaro}{cmyk}{1,.3,0.,0.1}
\hypersetup{colorlinks, citecolor=bluscuro, linkcolor=bluscuro, urlcolor=bluscuro}

\makeatother   
\newcommand{\GeV}{{\rm \,GeV}}
\newcommand{\TeV}{{\rm TeV}}

\newcommand{\met}{\slashed{E}_T}
\def\Qtr{Q_{\rm tr}}
\def\pT{p_{\rm T}}

\def\de{\textrm{d}}

 \def\be   {\begin{equation}}   \def\ee   {\end{equation}}
 \def\ba   {\begin{array}}      \def\ea   {\end{array}}
 \def\bea  {\begin{eqnarray}}   \def\eea  {\end{eqnarray}}
 \def\bean {\begin{eqnarray*}}  \def\eean {\end{eqnarray*}}
 \def\nn{\nonumber}

\begin{document}

\begin{flushright} 
SISSA  25/2014/FISI
\end{flushright}

\vspace{0.5cm}
\begin{center}

{\LARGE \textbf {
On the Validity of the Effective Field Theory
\\
[0.01cm]
 for Dark Matter Searches at the LHC
 \\
 [0.3cm]
 Part III:   Analysis for the $t$-channel
}}
\\ [1.5cm]

{\large
\textsc{Giorgio Busoni}$^{\rm a,}$\footnote{\texttt{giorgio.busoni@sissa.it}},
\textsc{Andrea De Simone}$^{\rm a, }$\footnote{\texttt{andrea.desimone@sissa.it}},
}
\\[0.2cm]
{\large
\textsc{Thomas Jacques}$^{\rm  b, }$\footnote{\texttt{thomas.jacques@unige.ch}},
\textsc{Enrico Morgante}$^{\rm b, }$\footnote{\texttt{enrico.morgante@unige.ch}},
\textsc{Antonio Riotto}$^{\rm b, }$\footnote{\texttt{antonio.riotto@unige.ch}}
}
\\[1cm]

\large{
$^{\rm a}$ 
\textit{SISSA and INFN, Sezione di Trieste, via Bonomea 265, I-34136 Trieste, Italy}\\
\vspace{1.5mm}
$^{b}$ 
\textit{Section de Physique, Universit\'e de Gen\`eve,\\
24 quai E. Ansermet, CH-1211 Geneva, Switzerland}
}
\end{center}

\vspace{0.5cm}

\begin{center}
\textbf{Abstract}
\begin{quote}
We extend our recent analysis of the limitations of the effective field theory approach to studying dark matter at the LHC, by investigating the case in which Dirac dark matter couples to standard model quarks via $t$-channel exchange of a heavy scalar mediator.
We provide analytical results for the validity of the effective field theory  description, for both $\sqrt{s}$ = 8  TeV and 14  TeV. We  make use of a MonteCarlo event generator  to assess the validity of our analytical conclusions.
We also point out the general trend that in the regions where the effective field theory is valid, the dark matter relic abundance is typically large.
\end{quote}
\end{center}

\def\thefootnote{\arabic{footnote}}
\setcounter{footnote}{0}
\pagestyle{empty}

\newpage
\pagestyle{plain}
\setcounter{page}{1}

\section{Introduction}

Despite overwhelming gravitational evidence for the existence of dark matter (DM), we still have very little information about its particle properties. Yet there is enough evidence to motivate a search for DM with a mass at the electroweak energy scale, with non-zero albeit very weak interactions with the Standard Model (SM), known as a weakly interacting massive particle (WIMP). Both direct and indirect detection  have been very successful at placing strong, model independent constraints on the WIMP-nucleon scattering rate and self-annihilation rate respectively \cite{Baudis:2012ig,Cirelli:2012tf,Feng:2014uja,Abramowski:2011hc,Ackermann:2013yva,Aprile:2012nq,Akerib:2013tjd}, and whilst there are anomalies that may be consistent with a WIMP signal \cite{Bernabei:2010mq,Agnese:2013rvf,Aalseth:2012if}, a conclusive discovery has not been achieved.
 
The LHC is searching for direct DM production at unprecedented energies, and has excellent potential to finally discover DM. 
 Mono-jet \cite{monojetATLAS1, monojetCMS1, monojetATLAS2, monojetCMS2}, mono-$W$/$Z$
\cite{ATLASWZ,Aad:2014vka}
and mono-photon \cite{monogammaATLAS1, monogammaCMS1, monogammaATLAS2, monogammaCMS2} searches  are currently  under way 
to look for an indirect signature of DM production.
Yet, given that the true nature of DM is unknown, it has proven difficult to constrain the WIMP sector as a whole in a model-independent way. One potential solution to this problem is the use of Effective Field Theories (EFTs), which allow a DM-SM interaction term to be written as a single effective operator, integrating out the mediator.\footnote{See e.g. Ref.~\cite{deSimone:2014pda,Alves:2013tqa} for recently proposed directions alternative to EFT and simplified models for DM searches at the LHC.} This has the advantage of reducing the parameter space to a single energy scale, $\Lambda$ (sometimes called $M_*$ in the literature), in addition to the DM mass, and reducing the potential number of WIMP models down to a relatively small basis set.

EFTs are inherently an approximation to a full UV-complete theory, and hence must be used with caution. Given that the LHC is operating at very large energies, it is important to ensure that constraints on EFTs are internally consistent and fall in a region where the EFT approximation is valid. 

This issue  has been investigated in Refs.  \cite{Busoni:2013lha,Busoni:2014sya} where the validity of the EFT at both $\sqrt{s}=$8 and 14 TeV has been tested when heavy mediators are exchanged in  the $s$-channel. In particular, the validity of the EFT was assessed by introducing a few quantities, some of them independent of the ultraviolet completion of the DM theory, which quantify the error made when using effective operators to describe processes with very high momentum transfer. 
It was found that only a small fraction of events were at energies where the EFT approximation is valid, regardless of the choice of cuts or operator. 
In addition, Refs.~\cite{Fox:2011pm,Buchmueller:2013dya} have compared constraints on some EFTs to those on simplified models where the mediator has not been integrated out, and found that constraints on $\Lambda$ using UV complete models can either be substantially stronger or substantially weaker than those constructed using EFTs, depending on the choice of parameters. Since the initial motivation of using EFTs is to place model independent constraints on the dark sector independent of assumptions about the input parameters, it is becoming clear that extreme caution must be used  when placing constraints on DM using EFTs at the LHC.

In this paper we extend the analysis of Refs.~\cite{Busoni:2013lha,Busoni:2014sya} to the $t$-channel. We consider a model where Dirac DM couples to SM quarks via $t$-channel exchange of a scalar mediator. The details of the model are described in section~\ref{sec:validity}.
Our goal is to determine in what regions of parameter space the EFT approach is a valid description of this model. The EFT approximation is made by integrating out the mediator particle, and combining the mediator mass $M$ with the coupling strength $g$ into a single energy scale, $\Lambda\equiv M/g$. This is done by expanding the propagator term for the mediator in powers of $Q_{\rm tr}^2/M^2$ and truncating at the lowest order, where $\Qtr$ is the momentum carried by the mediator:
\begin{align}
\frac{g^2}{Q_{\rm tr}^2-M^2}&=-\frac{g^2}{M^2}\left(1+\frac{Q^2_{\rm tr}}{M^2}+ \mathcal{O} \left(\frac{Q^4_{\rm tr}}{M^4}\right)\right) \\
&\simeq -\frac{1}{\Lambda^2}.
\end{align}
Clearly, this approximation is only valid when $\Qtr^2 \ll M^2$; yet this condition is impossible to test precisely in the true EFT limit, since $M$ has been combined with $g$ to form $\Lambda$. Instead, an assumption about $g$ must be made, defeating one of the primary advantages of EFTs. This is unavoidable, since the LHC operates at energies high enough that violation of the EFT approximation is a real concern and must be tested, as has been seen in Refs.~\cite{Busoni:2013lha,Busoni:2014sya}.
There is no lower limit to the unknown coupling strength $g$,\footnote{Although if $g$ is particularly small, and the DM is a thermal relic, then DM will be overproduced in the early universe unless another annihilation channel is available.} meaning that regardless of the scale of $\Lambda$, it is always possible that  $M$ is small enough that the EFT approximation does not apply, and the constraint on $\Lambda$ is invalid. In other words, for all operators, constraints on $\Lambda$ will only be valid down to a certain value of $g$ if the EFT approach has been taken. 

On the other hand, the most optimistic choice is to assume that $g \simeq 4 \pi$, the maximum possible coupling strength such that the model still lies in the perturbative regime. This choice is discussed later in the text.
As a middle ground, we test whether the EFT approximation is valid for values of $g\gtrsim 1$, a natural scale for the coupling in the absence of any other information. In this case, the condition for the validity of the EFT approximation becomes

\be
\Qtr^2 \lesssim \Lambda^2,
\ee
which we will adopt  in the following to assess the validity of the use of EFT at LHC for DM searches. 

The paper is organized as follows. Section \ref{sec:validity} contains the bulk of our analytical results for both $\sqrt{s}$ = 8 and 14 TeV and a comparison with those obtained  using fully numerical simulations of the LHC events. Our discussion and conclusions are summarized in section 
\ref{sec:conclusions}. 

\section{Validity of the EFT: analytical approach}
\label{sec:validity}

\subsection{Operators and cross sections}

In this paper we  will consider the following effective operator describing the interactions
between  Dirac dark matter $\chi$ and left-handed quarks $q$
\begin{equation}
\label{eq:t-op}
\mathcal{O} =\frac{1}{\Lambda^2} \left( \bar \chi P_L q \right) \, \left( \bar q P_R \chi\right).
\end{equation}
Only the coupling between dark matter and the first generation of quarks is considered. Including couplings to the other generations of quarks requires fixing the relationships between the couplings and mediator masses for each generation, making such an analysis less general. 
In principle the dark matter can also couple to the right-handed quark singlet,  switching $P_R$ and $P_L$ in the above operator. The inclusion of both of these operators does not modify our results, even if the two terms have different coupling strengths.

\begin{figure}[t!]
\centering
\includegraphics[width=0.35\textwidth]{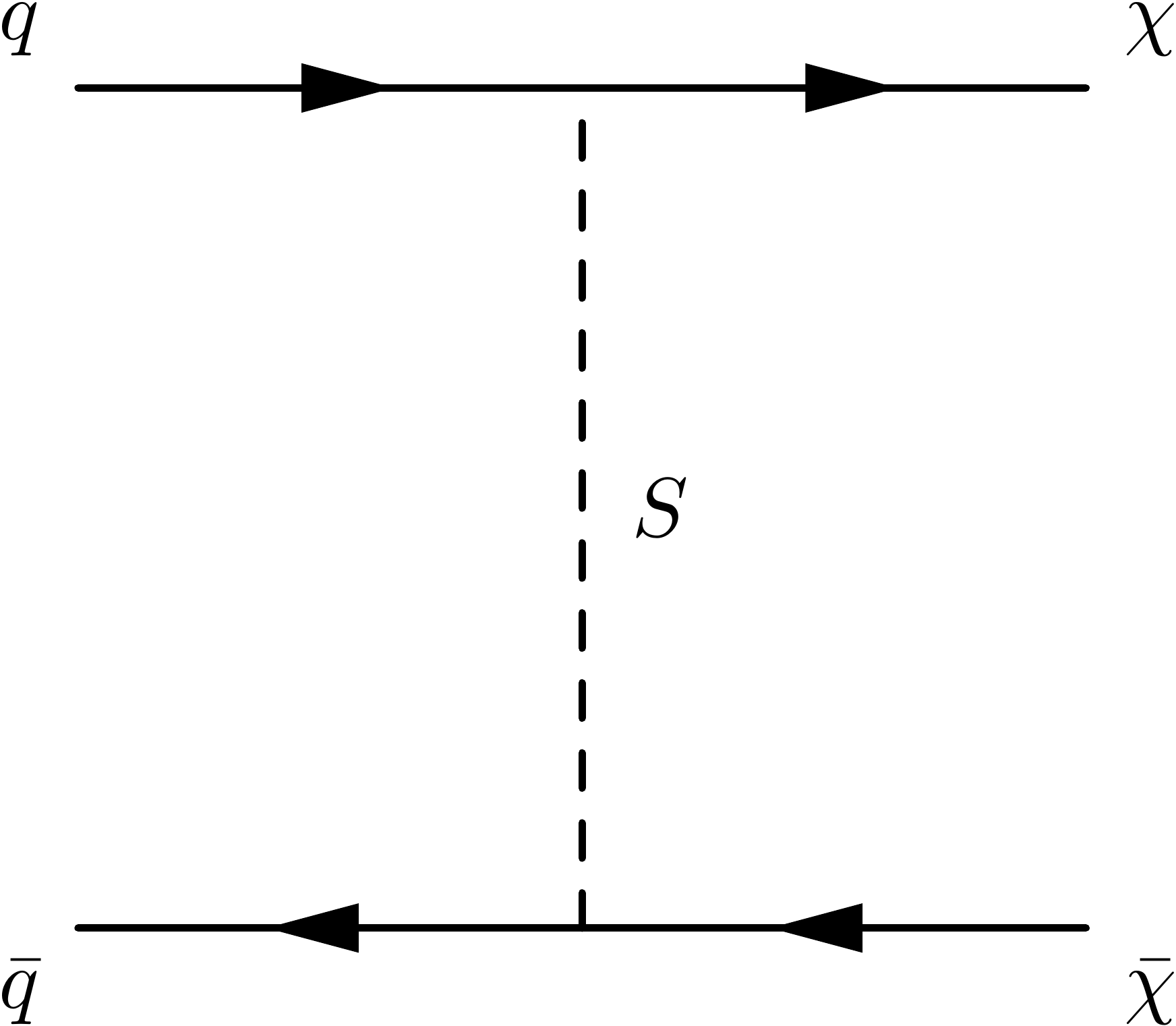}
\caption{
Illustrative $2\rightarrow 2$ process for the UV-complete version of our effective operator.}
\label{fig:diagrams}
\end{figure}

The operator in Eq.~(\ref{eq:t-op}) can be viewed as the low-energy limit of
a simplified model describing a quark doublet $Q_L$ coupling to DM, via $t$-channel exchange of a scalar mediator $S_Q$, 
\begin{equation}
\mathcal{L}_{\rm int} = g \,\bar \chi Q_L S_Q^* + h.c.
\label{eq:toy}
\end{equation}
and integrating out the mediator itself. Since we consider only coupling to the first generation of quarks, $Q_L = (u_L, d_L)$. 
%
As an illustration, the $2\rightarrow 2$ process $q\bar q\rightarrow \chi \bar \chi$ for this model is shown in Fig.~\ref{fig:diagrams}. This model is popular as an example of a simple DM model with $t$-channel couplings, which exist also in well-motivated models such as supersymmetry where the mediator particle is identified as a squark, and the DM is a Majorana particle. Bell et al. \cite{Bell:2012rg} have used a version of this model with Majorana DM in place of Dirac DM, to test the prospects of $Z$-bosons as a potential search channel. This has been followed up by a dedicated ATLAS search in this channel \cite{Aad:2014vka}. Refs \cite{Chang:2013oia,An:2013xka,Bai:2013iqa,DiFranzo:2013vra,Papucci:2014iwa,Garny:2014waa} have also constrained this model, using both the standard monojet search channel as well as searching for multiple jets arising from direct mediator production. Refs \cite{An:2013xka,Garny:2014waa} found that collider constraints on this model were competitive if not stronger than direct detection constraints across most of the parameter space.

The $t$-channel operator in Eqn.~(\ref{eq:t-op}) can be expressed as a sum of $s$-channel operators using Fierz transformations. For arbitrary Dirac spinors such as $\bar q_1,\,  q_2,\,  \bar\chi_1, \, \chi_2$, and adopting in part the notation of \cite{Bell:2010ei}, the Fierz transformation can be expressed as
\begin{equation}
\label{eq:fierz}
\left( \bar q_1 X \chi_2 \right) \left(\bar \chi_1 Y q_2\right) = \frac{1}{4} \sum_B \left( \bar q_1 X \Gamma^B Y q_2 \right) \left(\bar \chi_1 \Gamma_B \chi_2\right), 
\end{equation}
where $X$, $Y$ are some combination of Dirac-matrices, and $\Gamma^B = \{ \mathbb{1}, i \gamma_5, \gamma^\mu, \gamma_5 \gamma^\mu, \sigma^{\mu \nu} \}$ and $\Gamma_B = \{ \mathbb{1}, -i \gamma_5, \gamma_\mu, -\gamma_5 \gamma_\mu, \frac{1}{2}\sigma_{\mu \nu} \}$
form a basis spanning 4$\times$4 matrices over the complex number field \cite{Bell:2010ei}.
Due to the chiral coupling between the quarks and DM, most of the terms in the sum cancel, and we are left with
\begin{eqnarray}
\label{eq:fierz-op}
\mathcal{O} &=& \frac{1}{\Lambda^2} \left( \bar \chi P_L q \right) \, \left( \bar q P_R \chi\right) \nonumber\\
&=&\frac{1}{8\Lambda^2}\left( \bar \chi \gamma^\mu  \chi \right) \; \left( \bar q \gamma_\mu  q\right)\,\,\,\,\,\,\,\,\,\,\,\,\,(D5)\nonumber\\
&+&\frac{1}{8\Lambda^2}\left( \bar \chi \gamma^\mu\gamma_5  \chi \right) \; \left( \bar q \gamma_\mu  q\right)
\,\,\,\,\,\,\,\,(D6)\nonumber\\
&-&\frac{1}{8\Lambda^2}\left( \bar \chi \gamma^\mu  \chi \right) \; \left( \bar q \gamma_\mu\gamma_5  q\right)\,\,\,\,\,\,\,\,
(D7)\nonumber\\
&-&\frac{1}{8\Lambda^2}\left( \bar \chi \gamma^\mu\gamma_5  \chi \right) \; \left( \bar q \gamma_\mu\gamma_5  q\right)
\,\,\,(D8)\nonumber\\
&=& \frac{1}{2\Lambda^2} \left( \bar \chi \gamma^\mu P_R \chi \right) \; \left( \bar q \gamma_\mu P_L q\right).
\end{eqnarray}
This is equivalent to a rescaled sum of the  $D5$, $D6$, $D7$ and $D8$ operators \cite{Goodman:2010ku}.
Thus, it is interesting to see whether the EFT limit of the $t$-channel model under investigation has similar phenomenology to these $s$-channel operators. This is discussed in Section~\ref{sec:conclusions}.

The standard search channel for such a scenario is missing energy ($\met$) plus a single jet, although particles such as $Z$-bosons \cite{Bell:2012rg,Aad:2014vka}  are promising complementary search channels. The dijet$+ \met$ channel is particularly promising for the simplified model in Eq.~(\ref{eq:toy})   since direct production of a pair of mediator particles can result in a strong dijet signal. In particular, Refs. \cite{An:2013xka,Papucci:2014iwa} found that in much of parameter space, the dijet signal from direct mediator production provides comparable or stronger constraints on the model than the traditional monojet signal. 
In the high-energy limit, the mediator particle has SM charges and can emit a gluon, photon or massive gauge boson. This channel is suppressed in the EFT limit and so is not considered here.

\begin{figure}[t!]
\centering
\includegraphics[width=0.35\textwidth]{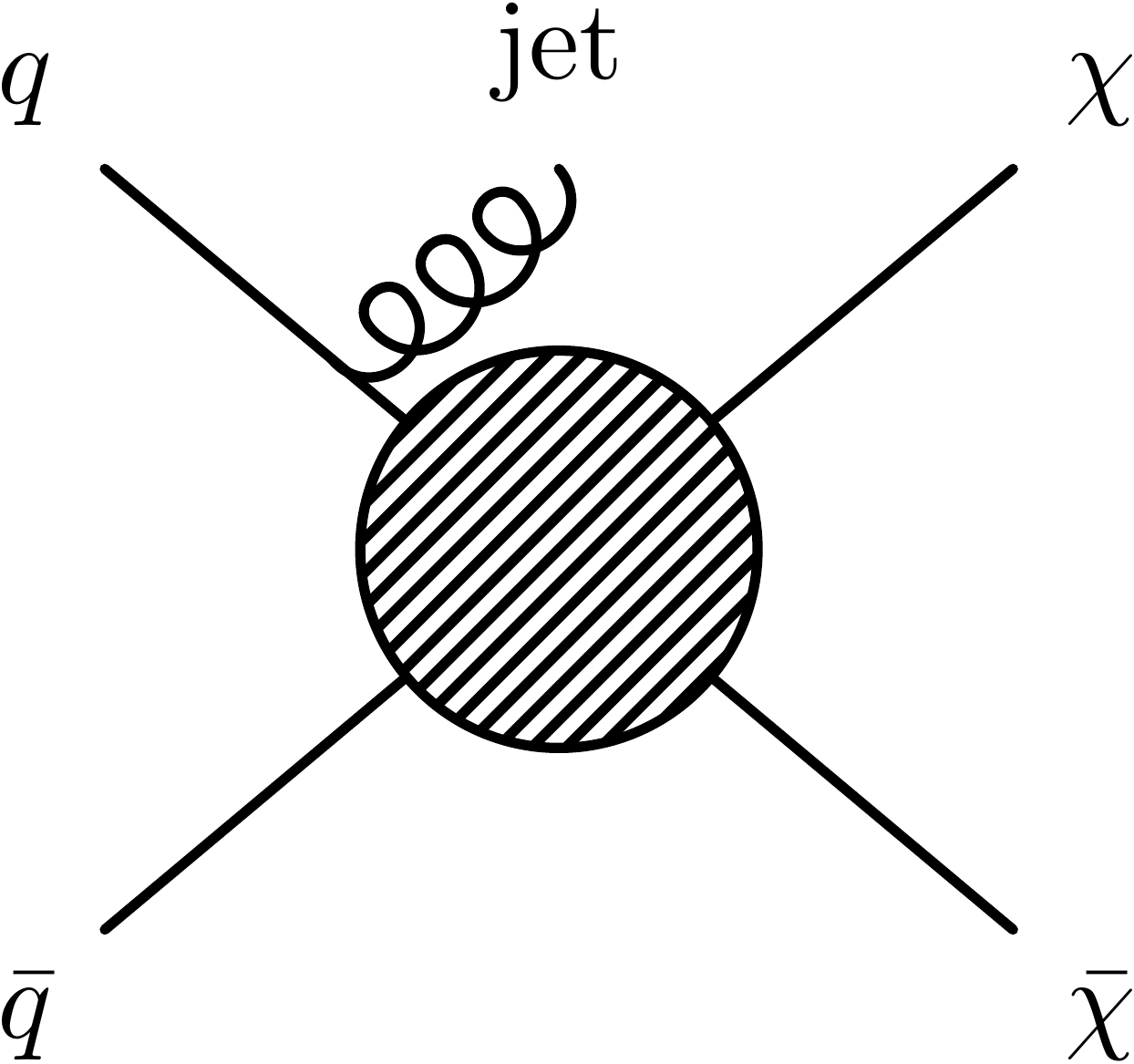}
\hspace{0.5cm} 
\includegraphics[width=0.35\textwidth]{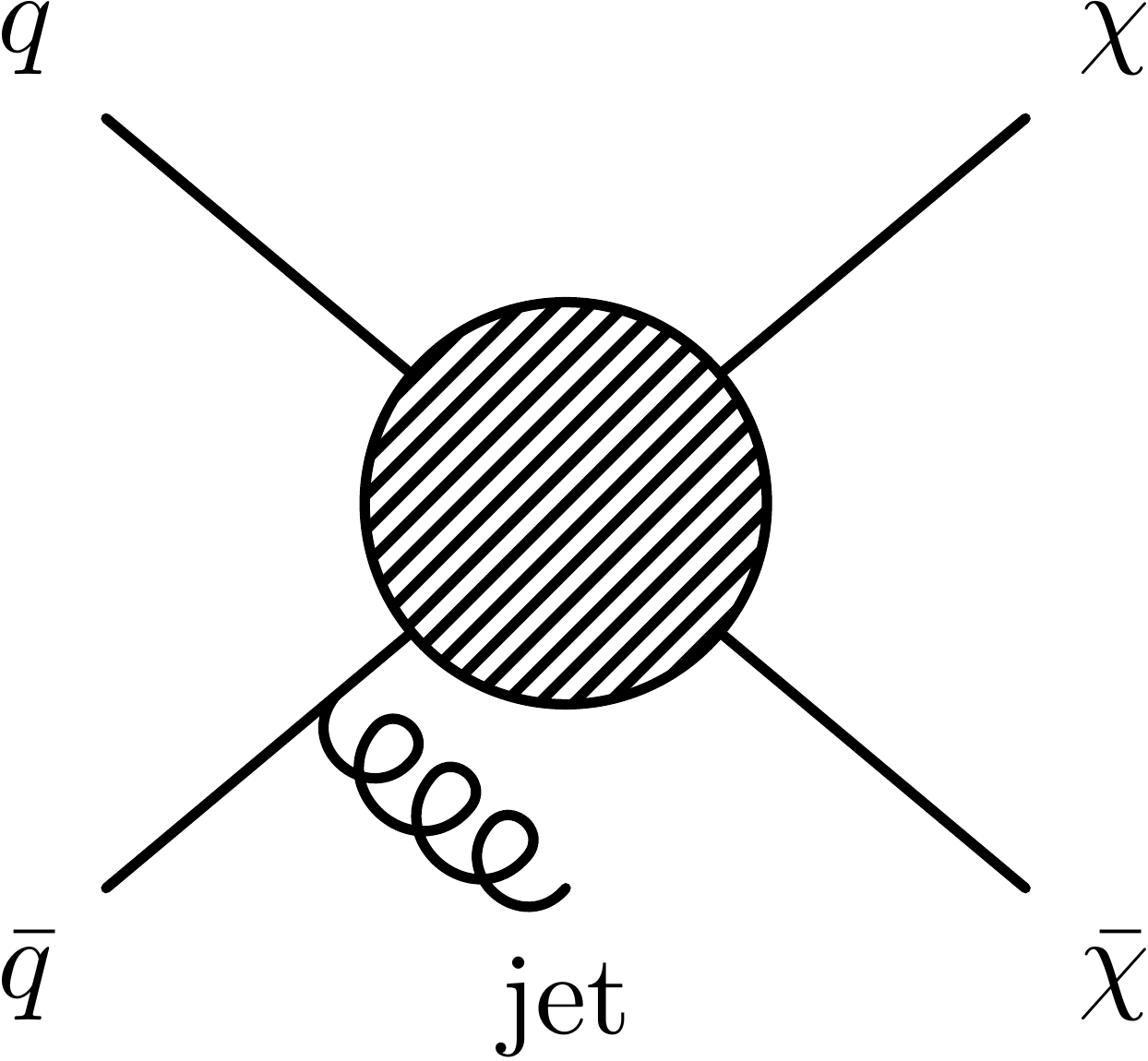}
\caption{
Search channel.
}
\label{fig:feynblob}
\end{figure}

The dominant process contributing to the $\met +$ monojet signal is $q \bar q \rightarrow \chi \bar \chi g$. Representations of the EFT diagrams are shown in Fig.~\ref{fig:feynblob}. We have calculated the differential cross section for these processes, with the results given in Appendix \ref{app:crosssect} along with the corresponding matrix elements. In the same appendix we have also calculated the differential cross section for the other contributing processes, $q g \rightarrow \chi \bar \chi q$ and $g \bar q \rightarrow \chi \bar  \chi \bar q$, which we found to be subdominant. In the full simplified model, the scalar mediator carries standard model charges and can emit gauge bosons, including gluons which would contribute to the  $\met +$ monojet signal. This channel is neglected in this study, since we are testing whether the effective operator description of this model is internally consistent regardless of the UV completion.

In order to compute the cross section with proton initial states appropriate for LHC events, it is necessary to integrate over the parton distribution function (PDF) of the proton. For   $q\bar q$ initial states, this is defined as
\be
\sigma=
\sum_q\int\de x_1\de x_2 
[f_q(x_1)f_{\bar q}(x_2)+f_q(x_2)f_{\bar q}(x_1)] \hat\sigma,
\ee
where $\hat\sigma$ is the total cross section for a process in the center of momentum frame. We have used the MSTW PDFs from Refs.~\cite{pdf1,pdf3}, and checked that the our results are not sensitive to the choice of leading or next-to-leading-order MSTW PDFs. 

\subsection{Results and discussion\label{sec:results}}

Recall that our goal is to determine whether the EFT approximation is valid for the operator in Eqn.~\ref{eq:t-op}, in the standard search channel $q \bar q \rightarrow \chi \bar \chi +$ jet, when the coupling strength is at roughly the natural scale, $1 \lesssim g \lesssim 4\pi$. In this case, for any given event, the momentum of the mediator can only be neglected if $\Qtr^2 \lesssim \Lambda^2$. To test this, we define the ratio of the cross section truncated so that all events pass the condition, to the total cross section: 

\be
R_\Lambda \equiv \frac{\sigma|_{\Qtr < \Lambda}}{\sigma} =
\frac{\int_{p_{\rm T}^{\rm min}}^{p_{\rm T}^{\rm max}}\de p_{\rm T}\int_{-2}^2\de \eta
\left.\dfrac{\de^2\sigma}{\de p_{\rm T}\de\eta}\right\vert_{Q_{\rm tr}<\Lambda}}
{\int_{p_{\rm T}^{\rm min}}^{p_{\rm T}^{\rm max}}\de p_{\rm T}\int_{-2}^2\de \eta
\dfrac{\de^2\sigma}{\de p_{\rm T}\de\eta}}.
\label{ratiolambdatot}
\ee
We have parameterised the cross section such that the final integration variables are the standard observables for jets observed at the LHC,  namely the transverse momentum $\pT$ and pseudorapidity $\eta$. The integration limits on these quantities are chosen to be comparable to those used in standard searches for WIMP DM by the LHC collaborations (see, for instance, Ref. \cite{monojetATLAS2}). For searches at center of mass energy $\sqrt{s} = 8$ TeV, $\pT$ is integrated from 500 GeV to 1 TeV. For $\sqrt{s} = 14$ TeV, the integration range is instead 500 GeV to 2 TeV. In both cases, the pseudorapidity integration range is $| \eta| \leq 2$. 

There are two values of $\Qtr$, corresponding to jet emission from either the initial state quark or antiquark respectively. 
These are given in Appendix \ref{app:Qtr}. Mixing between diagrams makes it impossible to disentangle a single transferred momentum for any individual event, and so we require that for each event \emph{both} values of $\Qtr$ for that process satisfy the requirement that $\Qtr^2 < \Lambda^2$. 

\begin{figure}[t!]
\centering
\includegraphics[width=0.45\textwidth]{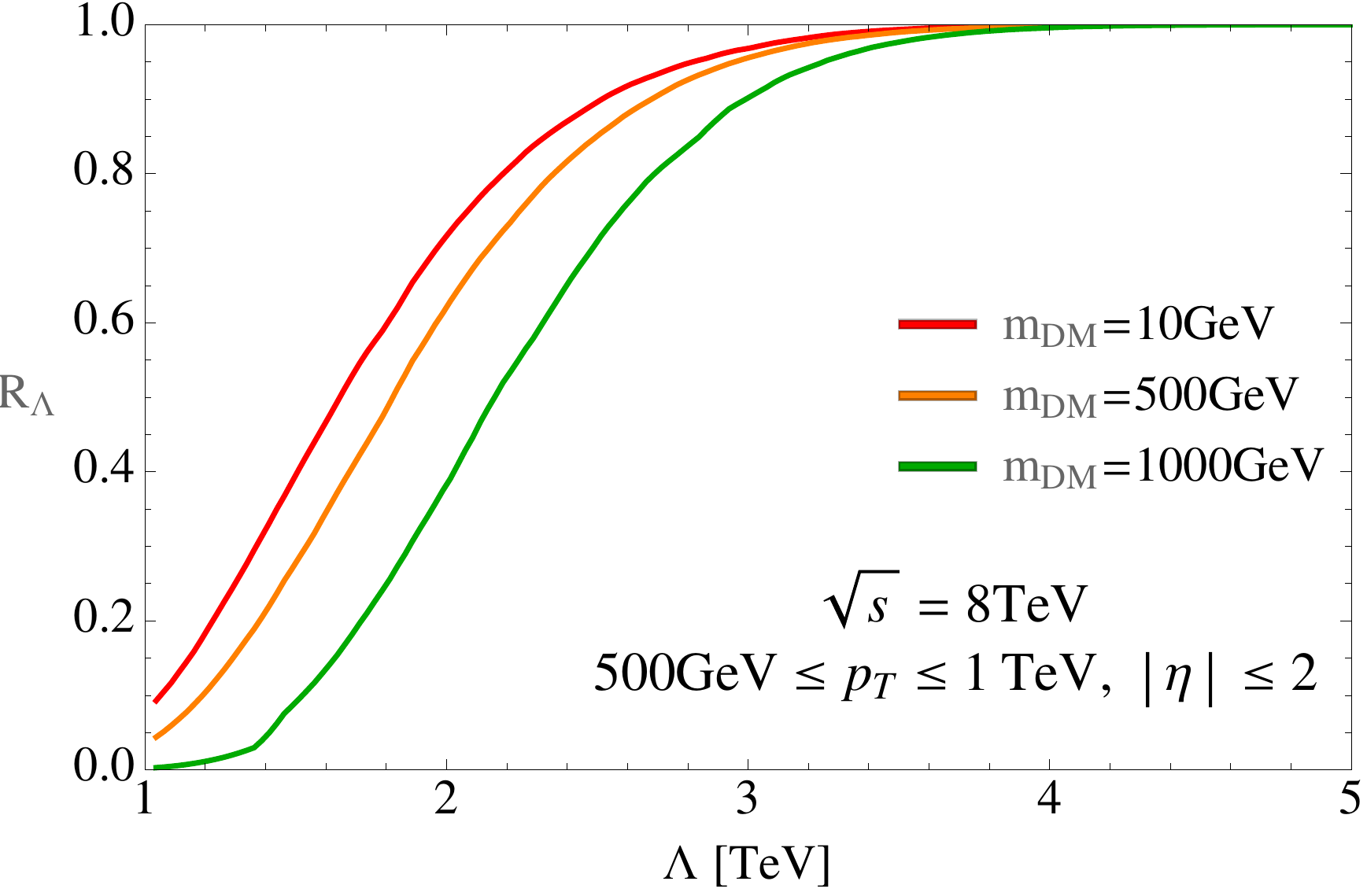}
\hspace{0.5cm} 
\includegraphics[width=0.45\textwidth]{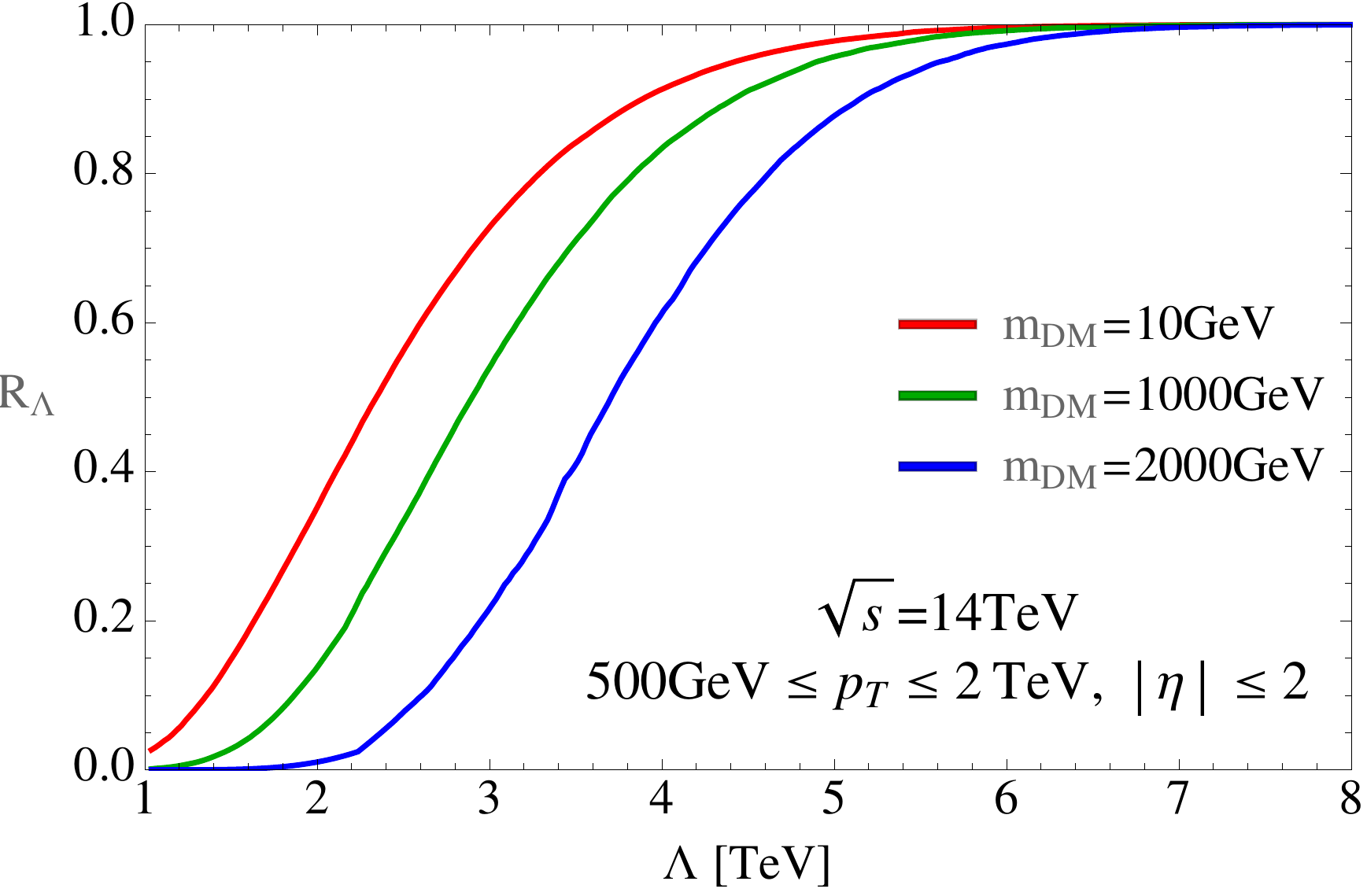}\\
\caption{
 The ratio $R_{\Lambda}^{\rm tot}$ as a function of $\Lambda$ for three choices of the DM mass,
  for $\sqrt{s}=8$ TeV (left panel) and $14$ TeV (right panel).
}
\label{fig:RvsL}
\end{figure}

\begin{figure}[t!]
\centering
\includegraphics[width=0.45\textwidth]{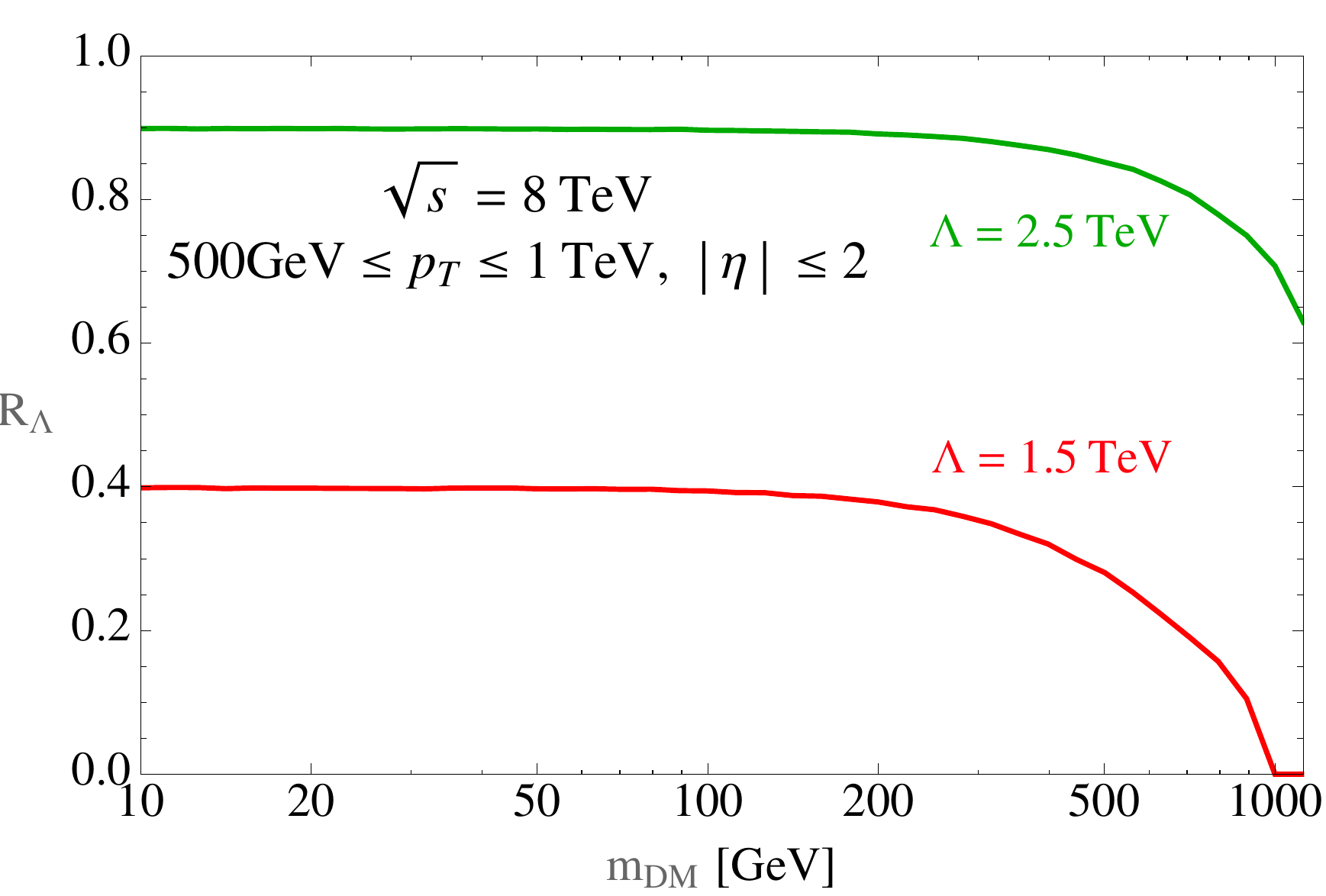}
\hspace{0.5cm} 
\includegraphics[width=0.45\textwidth]{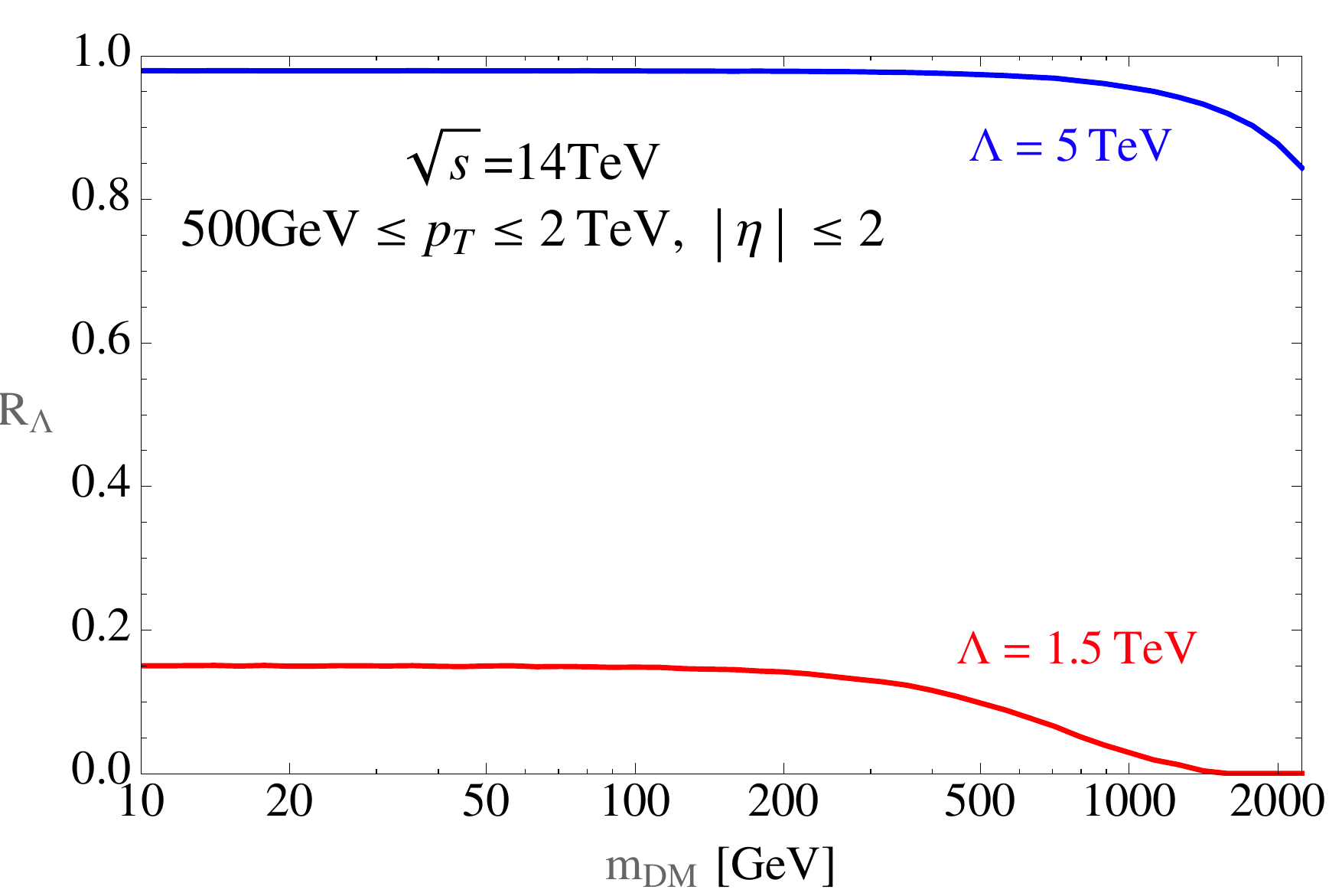}\\
\caption{
 The ratio $R_{\Lambda}^{\rm tot}$ as a function of $m_{\rm DM}$ for two choices of $\Lambda$,
  for $\sqrt{s}=8$ TeV (left panel) and $14$ TeV (right panel).
}
\label{fig:RvsMx}
\end{figure}

\begin{figure}[t!]
\centering
\includegraphics[width=0.45\textwidth]{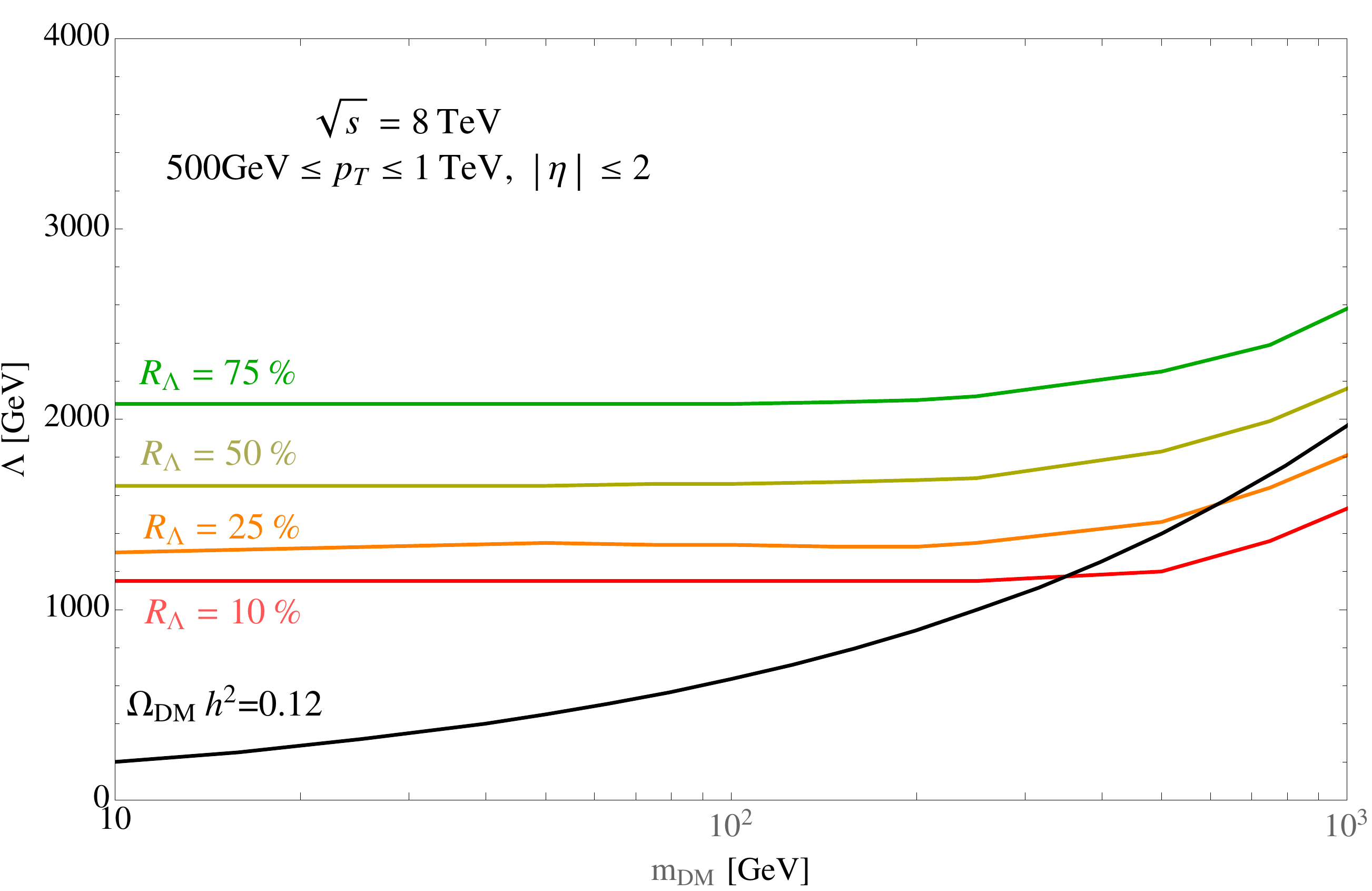} 
\hspace{0.5cm}
\includegraphics[width=0.45\textwidth]{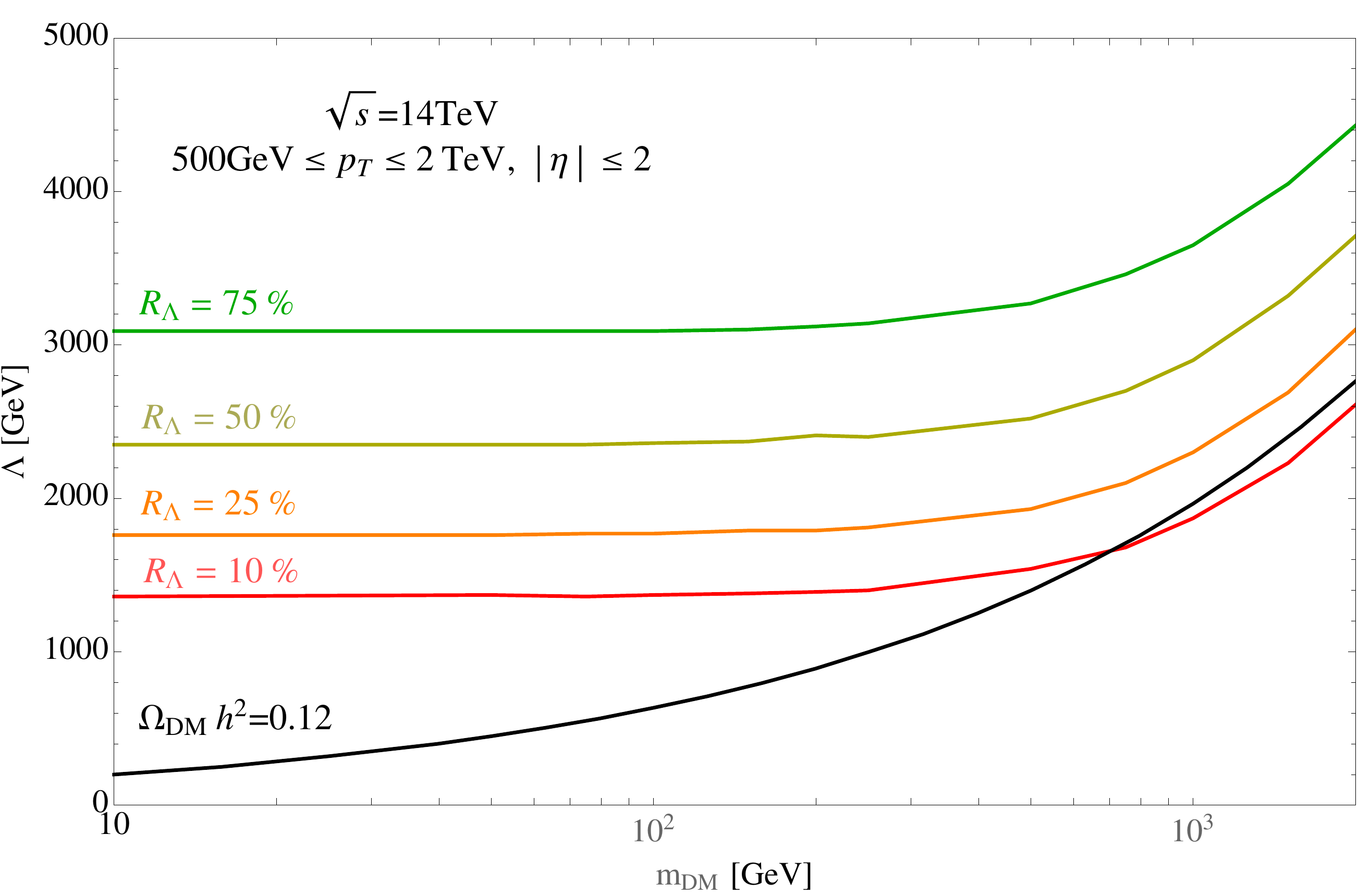} 
\caption{
Contours for the ratio $R_{\Lambda}^{\rm tot}$, defined in Eq.~(\ref{ratiolambdatot}), on the plane
$(m_{\rm DM}, \Lambda)$. We set $\sqrt{s}=8 {\rm TeV},
|\eta|\leq 2$ and $500 \GeV<p_{\rm T}<1 \,\TeV$ in the left panel, and $\sqrt{s}=14 {\rm TeV},
|\eta|\leq 2$ and $500 \GeV<p_{\rm T}<2 \,\TeV$ in the right panel. 
  The black solid curves indicates the correct relic abundance.
}
\label{fig:RContours1}
\end{figure}

\begin{figure}[t!]
\centering
\includegraphics[width=0.45\textwidth]{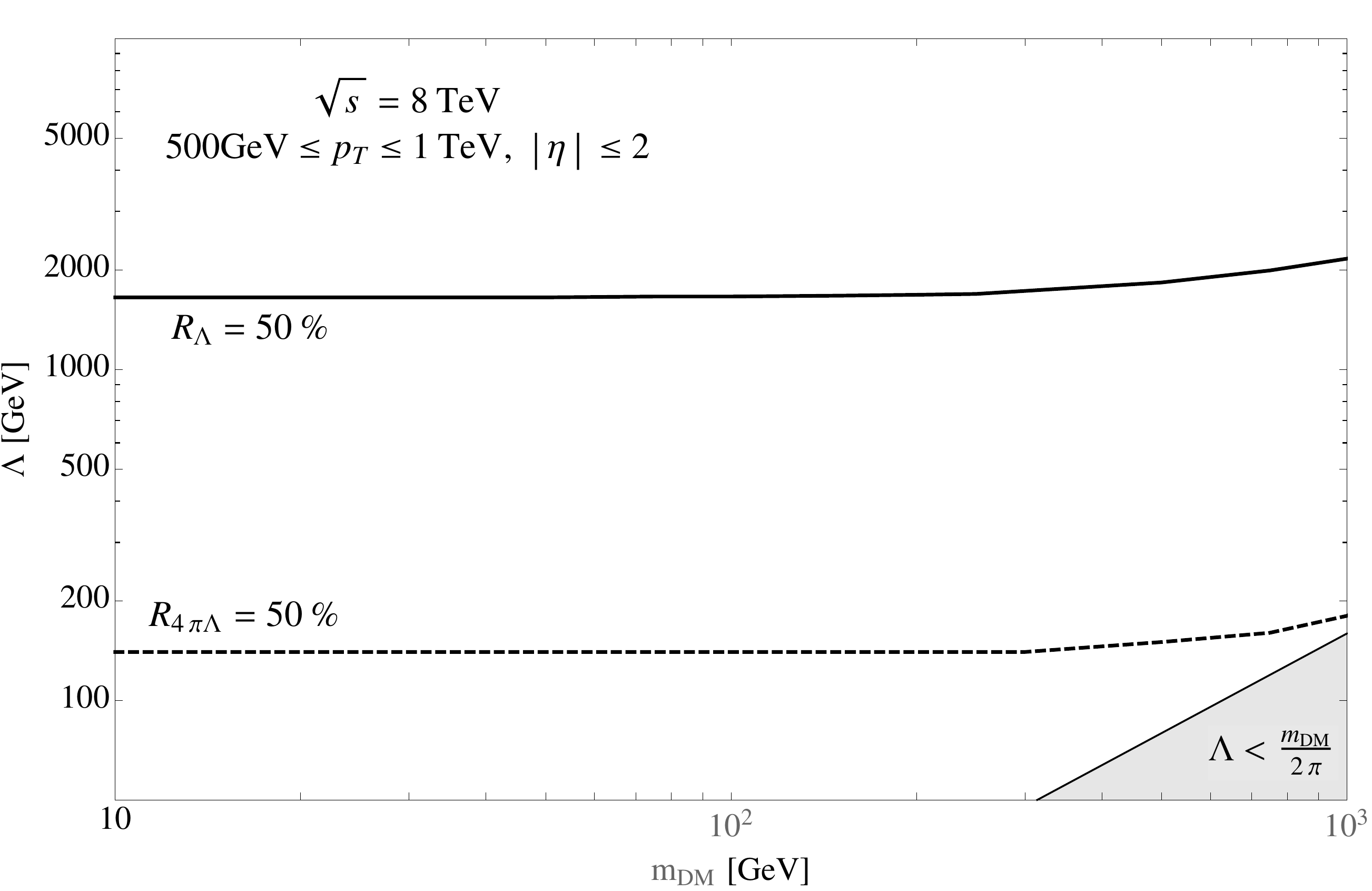} 
\hspace{0.5cm}
\includegraphics[width=0.45\textwidth]{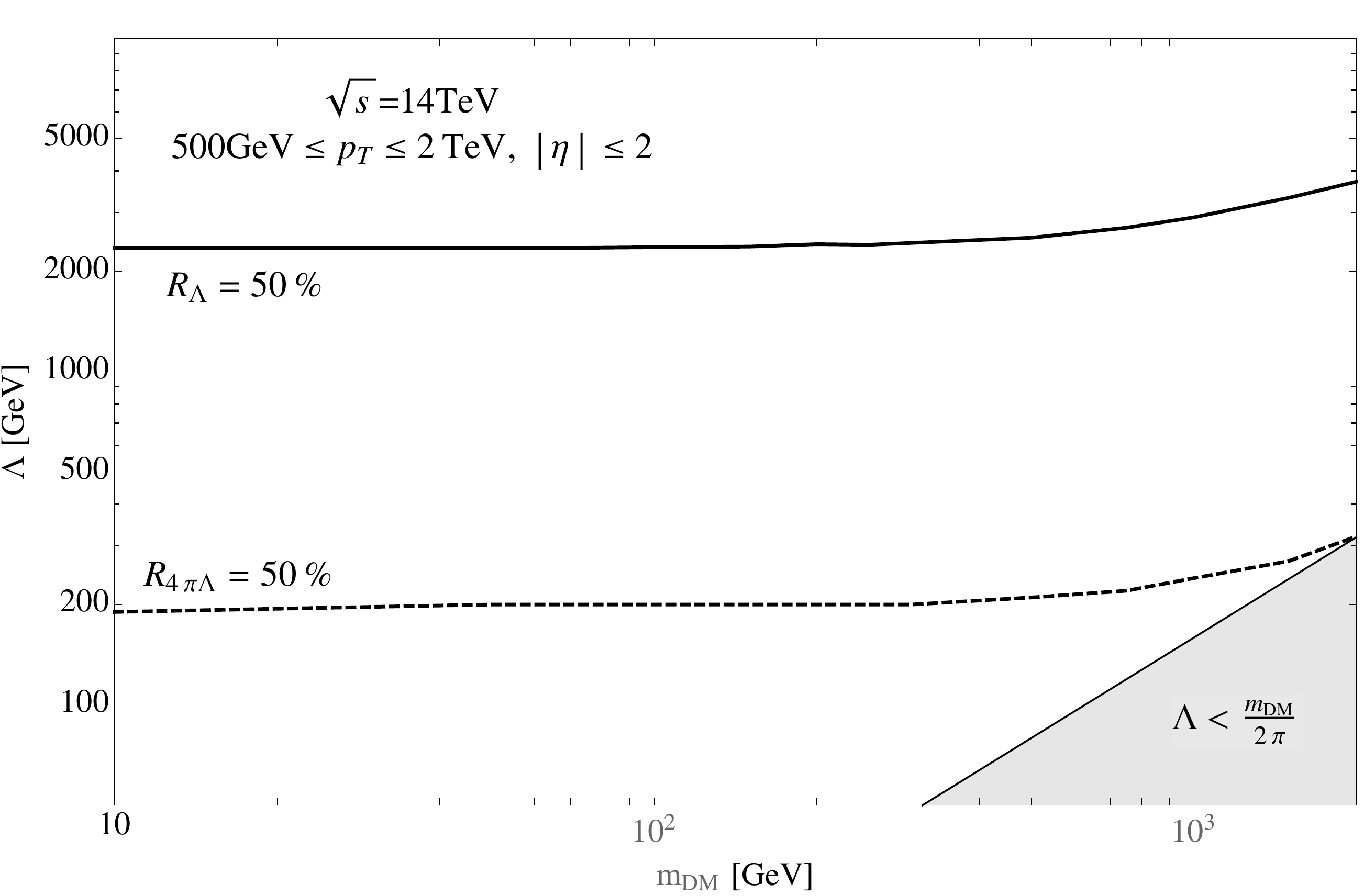}
\caption{
50\% contours for the ratio $R_{\Lambda}^{\rm tot}$, 
varying the cutoff $Q_{\rm tr}<\Lambda$ (solid line) and $Q_{\rm tr}<4\pi\Lambda$ (dot-dashed
line).
We have also shown the region corresponding to $\Lambda<m_{\rm DM}/(2\pi)$ (gray shaded area),
often used as a benchmark for the validity of the EFT.
 We set $\sqrt{s}=8$ {\rm TeV} (left panel) and $\sqrt{s}=14$ {\rm TeV} (right panel).
}
\label{fig:RContours2}
\end{figure}

In Fig.~\ref{fig:RvsL} we show the behaviour of $R_\Lambda$ as a function of $\Lambda$, at both $\sqrt{s}=$8 and 14 TeV. 
Similarly, Fig.~\ref{fig:RvsMx} shows $R_\Lambda$ as a function of $m_{\rm DM}$ at the same center of mass energies. 
%
In Fig.~\ref{fig:RContours1} we instead plot isocontours of four fixed values of $R_\Lambda$ as a function of both $m_{\rm DM}$ and $\Lambda$. 
 Contrasted with the $s$-channel case \cite{Busoni:2013lha,Busoni:2014sya}, the ratio has less DM mass dependence, being even smaller than in the $s$-channel case at low DM masses and larger at large DM masses, without becoming large enough to save EFTs.
%

In Fig.~\ref{fig:RContours1} we also show the curves corresponding to the correct DM relic density, 
assuming that interactions between the DM particle and the SM plasma were mediated by 
the operator (\ref{eq:t-op}). These were computed  by using a semi-analytic solution to the Boltzmann equation \cite{Bertone:2004pz} to find the values of $m_{\rm DM}$ and $\Lambda$ that yield a DM abundance matching the observed value $\Omega_{\rm DM} h^2 \simeq 0.12$ \cite{Ade:2013zuv}. Since we are dealing with Dirac DM, have included an additional factor of 2 in the expression for the relic density relative to the equation for Majorana DM in Ref.~\cite{Bertone:2004pz}. For given $m_{\rm DM}$, larger $\Lambda$ leads to a smaller
self-annihilation cross section and therefore to larger relic abundance. It is evident that the large-$\Lambda$ region
where the  EFT is valid typically leads to an unacceptably large DM density.
However, it may certainly be that additional annihilation channels and interactions,
 beyond those described by the operator (\ref{eq:t-op}) can enhance the cross section and
 decrease the relic abundance to fit the observations.

In the most optimistic scenario for EFTs, the coupling strength $g$ takes the maximum value  ($4\pi$) such that the model remains in the perturbative regime. In this case, a given constraint on $\Lambda$ corresponds to a relatively larger value of $M$, such that the EFT is valid across a larger region. To demonstrate how our results depend on the coupling strength, in Fig.~\ref{fig:RContours2} we plot isocontours for $R=50\%$, for two cases: {\bf1)} the standard requirement that $\Qtr^2 < \Lambda^2$, equivalent to requiring $g \simeq 1$, and {\bf 2)} requiring $\Qtr^2 < (4 \pi \Lambda)^2$, equivalent to requiring $g \simeq 4 \pi$.  

The grey shaded area indicates the region where $\Lambda < m_{\rm DM}/(2\pi)$. This is often used as a benchmark for the validity of the EFT approximation, since in the $s$-channel,  $\Qtr$ is kinematically forced to be greater than $2m_{\rm DM}$, leaving the  EFT inherently invalid when $M<2m_{\rm DM}$, which is equivalent to $\Lambda < m_{\rm DM}/(2\pi)$ for a coupling strength $g\simeq 4\pi$. Thus,
in the $s$-channel the contours never cross this boundary. Interestingly this is not the case in the $t$-channel, since the kinematic constraints on $\Qtr$ no longer apply. This indicates that at very large DM masses the EFT approximation can become safer than naively assumed - although in practice the ratio is still too low for EFTs to be of any practical use.

\begin{figure}[t!]
\centering
\includegraphics[width=0.45\textwidth]{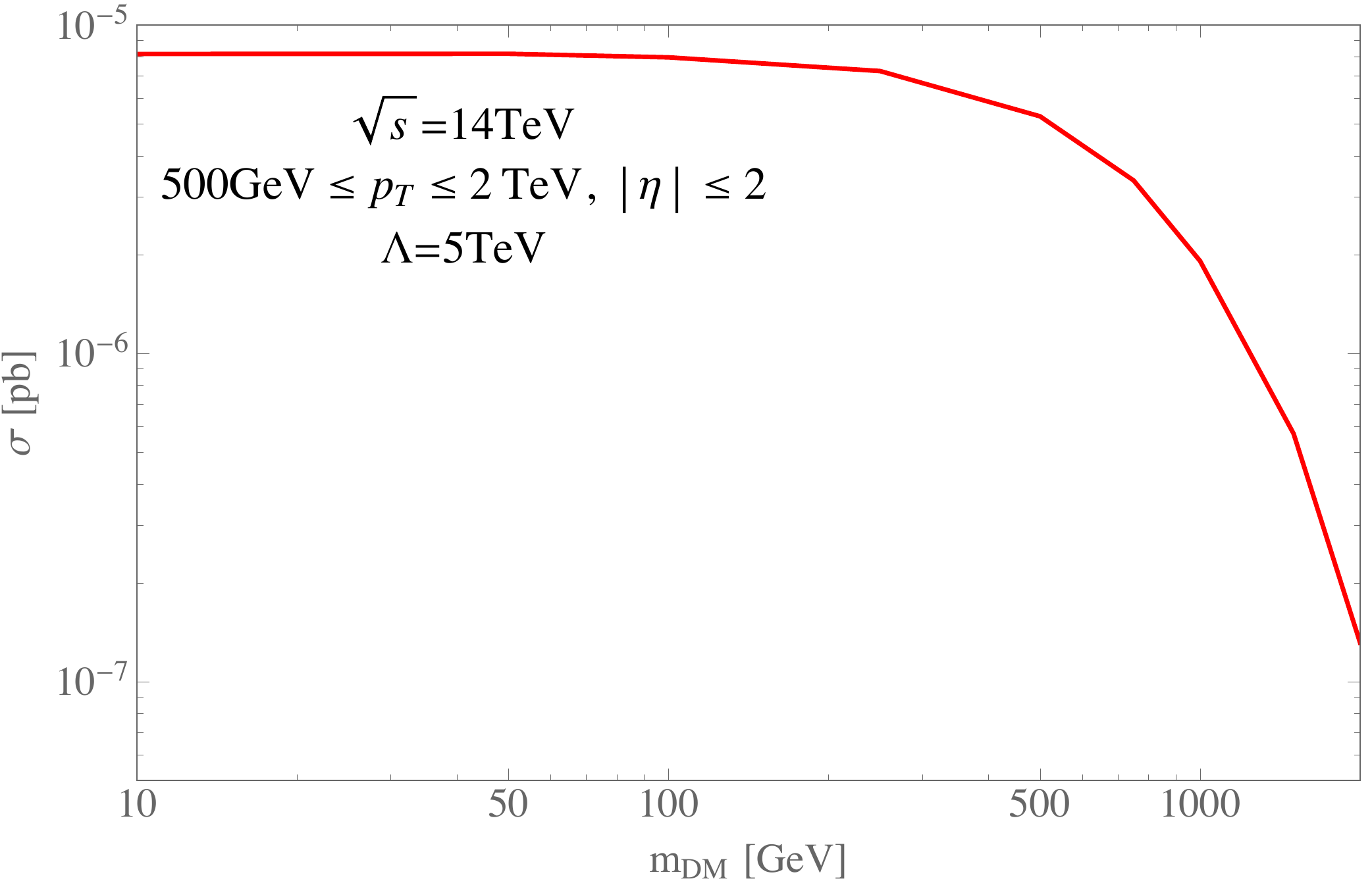}
\caption{
Cross section for the monojet process under consideration, before applying $R_\Lambda$ cuts. Note that $\sigma \propto \Lambda^{-4}$.}
\label{fig:rate}
\end{figure}

To gain a sense of whether this model is potentially observable at the LHC, and whether the effective operator model is still observable even after rescaling by $R_\Lambda$, we show in Figure~\ref{fig:rate} the integrated signal cross-section at $\sqrt{s} = 14$ TeV, using the same cuts as earlier. We can see even at relatively low dark matter masses, $\Lambda$ must be smaller than $\sim$1 TeV before events can be expected to be produced after 25 fb$^{-1}$, at which point the effective operator approach has entirely ceased to be a valid approximation. At higher luminosities the model will begin to become more observable for a greater range of $\Lambda$.

\subsection{Comparison with MonteCarlo Simulations\label{sec:numerical}}

As a check, it is interesting to compare our analytical results to fully numerical results. We have reproduced Fig.~\ref{fig:RvsL} using numerical  simulations of the LHC events at truth level, i.e., simulating events as they would be produced in truth without simulating how they would be observed by the ATLAS or CMS detectors.

The $t$-channel EFT model from Eqn.~\ref{eq:t-op} was constructed using FeynRules \cite{Christensen:2008py}, and the resultant Feynman rules were exported into \textsc{MadGraph 5}\cite{mg5}. The process of interest, $p p \rightarrow \chi \bar\chi+$ jet, was simulated at a center-of-mass energy of $\sqrt{s}=$ 14 TeV using the CTEQ6L1 PDF set \cite{Pumplin:2002vw}. It was found in Ref.~\cite{Busoni:2014sya} that the choice of PDF influences the magnitude but not the acceptance of the rate, and therefore this different choice of PDF relative to our analytic calculations is not expected to influence the ratios we calculate.
 Contours in $R_\Lambda$, defined in the same way as in Section~\ref{sec:results}, were determined by counting the fraction of events that passed the condition $\Qtr < \Lambda$, for both values of $\Qtr$ defined in Appendix~\ref{app:Qtr}. Events were simulated at DM masses of 10, 50, 100, 200, 300, 500, 1000 and 2000 GeV for a wide range of values of the cutoff scale $\Lambda$. The transverse momentum and rapidity of the jet are restricted to the ranges $(500 \leq p_{\rm T}/{\rm GeV} \leq 2000)$ and $|\eta| \leq 2$ respectively, as in the analytic results from the previous section.

\begin{figure}[t!]
\centering
\includegraphics[width=0.55\textwidth]{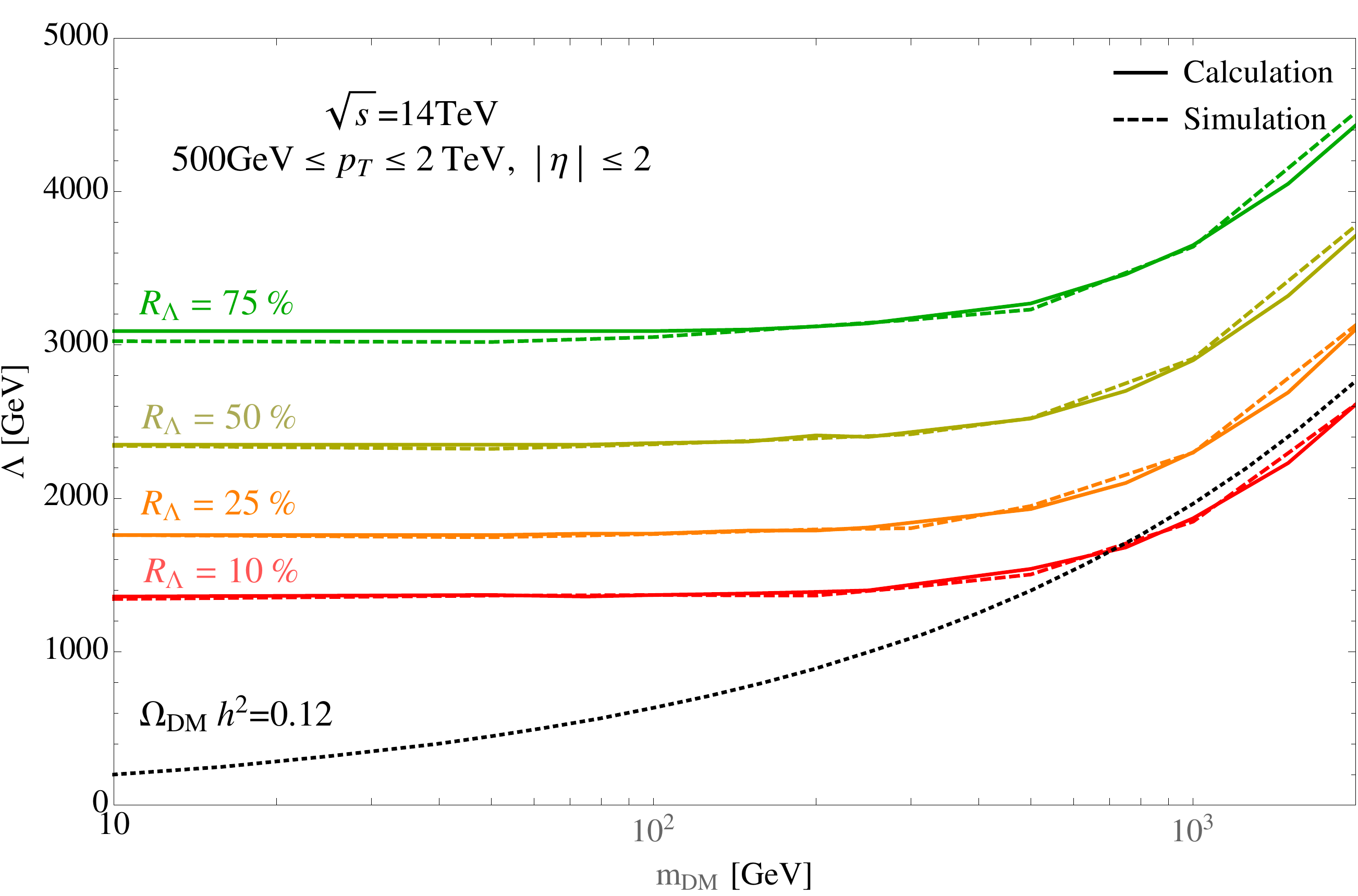}
\caption{
Comparison of the contour $R_\Lambda^{\rm tot} = 50 \%$ for the analytical calculation (solid line) and the simulation (dashed line). The dotted curve indicates the correct relic abundance.
}
\label{fig:calcsimcomp}
\end{figure}

\section{Conclusions}
\label{sec:conclusions}
In this paper we have extended  the investigation of the validity of the EFT approach for DM searches
at the LHC. While in Refs. \cite{Busoni:2013lha,Busoni:2014sya} our analysis has been focused on the
case of the EFT operators generated by integrating out a heavy mediator in the $s$-channel, we have 
considered here the case of Dirac DM couplings to the standard model via the $t$-channel. Even though  a $t$-channel operator can be expressed by a Fierz 
transformation as a sum of $s$-channel operators, our  results as a function of $\Lambda$ and DM mass (compared, for instance, to those of Fig. 2 of Ref. \cite{Busoni:2014sya}) indicate that one may not infer them from  those of a single $s$-channel operator,  see Eq. (\ref{eq:fierz-op}). This is  due to the
inherently different kinematics of the $s$- and $t$-channel, in particular significant differences in the transferred momentum.

We have also computed the relic density  over the parameter space of the model,
assuming that the only interactions between DM and the SM are those mediated by the $t$-channel operator (\ref{eq:t-op}), and found that the region of EFT validity corresponds to an overly large relic density.
This conclusion is rather general and may be evaded by assuming additional DM 
annihilation channels.

Similar to what happens in the $s$-channel case, our findings indicate that in the $t$-channel the range of validity of the EFT is 
significantly limited in
the parameter space $(\Lambda,m_{\rm DM})$, reinforcing the need to go beyond the
EFT at the LHC when looking for DM signals. This is especially true for light mediators as they can  be singly produced in association  with
a DM particle, leading to a qualitatively new
contribution to the mono-jet processes. Mediators  can
 even be pair-produced at the LHC through both QCD processes
and  DM exchange processes. All of this rich dynamics leads to stronger signals (and therefore, in the absence thereof, to tighter bounds) than the EFT approach. 

\section*{Acknowledgments}
We thank A. Brennan, C. Doglioni, G. Iacobucci, S. Schramm and S. Vallecorsa for many interesting conversations.
ADS acknowledges partial support from the  European Union FP7  ITN INVISIBLES 
(Marie Curie Actions, PITN-GA-2011-289442).

\appendix
\section{Three-body Cross Sections\label{app:crosssect}}

In this Appendix we show the details of the calculations of the tree-level cross sections for
the hard scattering processes $q(p_1)+\bar q(p_2)\to \chi(p_3)+\chi(p_4)+g(k)$ and $q(p_1)+g(p_2)\to \chi(p_3)+\chi(p_4)+q(k)$,
computed using the effective Lagrangian of Eq.~\ref{eq:t-op}.

\subsection{Matrix Elements\label{app:elements}}
The amplitudes for the process we are interested in are described, at leading order, by the Feyman diagrams in figure \ref{fig:diagrams}.
In the EFT limit they are given by
\bea
{\cal M}_{g}&=&- i \, \frac{g^2 g_s}{M^2} \epsilon_\mu^* T^a_{ij} \times \nn\\
&&\hspace{-.5cm} \times \left\{
\frac{\bar{u}(p_3)P_L(\cancel p_1 -\cancel p_2)\gamma^\mu u(p_1) \bar{v}(p_2)P_R v(p_4)}{(p_1-k)^2}  - 
\frac{\bar{u}(p_3)P_L u(p_1) \bar{v}(p_2)P_R \gamma^\mu (\cancel p_2 - \cancel k) v(p_4)}{(p_2-k)^2}
\right\} \nn\\
{\cal M}_{q}&=&- i \, \frac{g^2 g_s}{M^2} \epsilon_\mu T^a_{ij} \times \nn\\
&&\hspace{-.5cm} \times \left\{
\frac{\bar{u}(k)P_R v(p_3) \bar{u}(p_4)P_L(\cancel p_1 + \cancel p_2) \gamma^\mu u(p_1)}{(p_1+p_2)^2}  - 
\frac{\bar{u}(k)\gamma^\mu(\cancel p_2 - \cancel k)P_R v(p_3) \bar{u}(p_4) P_L u(p_1)}{(p_2-k)^2}
\right\} \nn\\
{\cal M}_{\bar q}&=&- i \, \frac{g^2 g_s}{M^2} \epsilon_\mu T^a_{ij} \times \nn\\
&&\hspace{-.5cm} \times \left\{
\frac{\bar{v}(k)P_L u(p_3) \bar{v}(p_4)P_R(\cancel p_1 + \cancel p_2) \gamma^\mu v(p_1)}{(p_1+p_2)^2}  - 
\frac{\bar{v}(k)\gamma^\mu(\cancel p_2 - \cancel k)P_L u(p_3) \bar{v}(p_4) P_R v(p_1)}{(p_2-k)^2}
\right\} 
\eea
for the gluon, quark and anti-quark emission processes respectively. Here we denote  the gluon polarization vector by $\epsilon_\mu$
and  the left and right projectors  $(1-\gamma_5)/2$ and $(1+\gamma_5)/2$  with $P_L$ and $P_R$ respectively.
The matrix $T^a_{ij}$ stands for the standard QCD Gell-Mann matrices. Notice that we work in the massless quark limit.
The anti-quark matrix element is simply obtained from the quark one by exchanging quarks with anti-quarks and left with right projectors.
The parton level cross sections for the two processes are thus the same, so here we only show the explicit derivation of the quark one.
The squared amplitudes, averaged over initial states (spin and colour) and summed over the final states, are given by
\bea\label{eq:Msqavgluon}
\left\vert{\cal M}_{g}\right\vert^2 & = & \frac{1}{9}\frac{g_s^2}{\Lambda^4}\frac{1}{(k\cdot p_1)(k\cdot p_2)} \times \nn\\
&&\hspace{-0.5cm}
\bigg\{
p_1\cdot p_3 \Big[ (k\cdot p_4)(k\cdot p_1) - (k\cdot p_4)(p_1\cdot p_2) - (k\cdot p_2)(p_1\cdot p_4) \Big] + \nn\\
&&\hspace{-0.5cm} +
p_2\cdot p_4 \Big[ (k\cdot p_3)(k\cdot p_2) - (k\cdot p_3)(p_1\cdot p_2) - (k\cdot p_1)(p_2\cdot p_3) \Big] + \nn\\
&&\hspace{-0.5cm} +
 (p_1\cdot p_3)(p_2\cdot p_4) \Big[ 2p_1\cdot p_2 -k\cdot p_1 -k\cdot p_2 \Big]
\bigg\}
\eea
\bea\label{eq:Msqavquark}
\left\vert{\cal M}_{q}\right\vert^2 & = & \frac{1}{6}\frac{g_s^2}{\Lambda^4}\frac{1}{(k\cdot p_1)(k\cdot p_2)} \times \nn\\
&&\hspace{-0.5cm}
\bigg\{
p_1\cdot p_4 \Big[ (k\cdot p_2)(p_1\cdot p_3) - (k\cdot p_1)(p_2\cdot p_3) +  \nn\\
&&\hspace{.5cm}+ (k\cdot p_2)(k\cdot p_3) + (k\cdot p_1)(k\cdot p_3) - (p_1 \cdot p_2)(k\cdot p_3) \Big] + \nn\\
&&\hspace{-0.5cm} +
p_2\cdot p_2 \Big[ (p_1\cdot p_4)(p_2\cdot p_3) - (k\cdot p_4)(k\cdot p_3) \Big] + \nn\\
&&\hspace{-0.5cm} +
 (k\cdot p_3) \Big[ (k\cdot p_1)(p_1\cdot p_4) + (k\cdot p_1)(p_2\cdot p_4) + (k\cdot p_2)(p_2\cdot p_4) \Big]
\bigg\}
\eea
with $\Lambda^2 = M^2/g^2$. 

\subsection{Cross sections\label{app:sigma}}

The simplest way to compute the cross section in the lab frame is to first evaluate the matrix elements and the phase space density
in the center-of-mass (c.o.m.)  frame, and then boost the result to the lab frame. 
In the c.o.m. frame, let us parametrize the four-momenta inolved in the process
as
\bea
\label{eq:Param}
p_1&=&x \frac{\sqrt{s}}{2}(1,0,0,1)\,,\qquad 
p_2=x \frac{\sqrt{s}}{2}(1,0,0,-1)\,, \qquad
k= x\frac{\sqrt{s}}{2}(z_0,z_0\hat k)\, ,\\
p_3&=&x \frac{\sqrt{s}}{2}(1-y_0,\sqrt{(1-y_0)^2-a^2}\hat p_3 )\,,\quad
p_4=x \frac{\sqrt{s}}{2}(1+y_0-z_0,\sqrt{(1+y_0-z_0)^2-a^2}\hat p_4)\, ,
\nn
\eea
where the two colliding partons  carry equal momentum fractions $x_1=x_2\equiv x$ of the incoming
protons, $a\equiv 2 m_{\rm DM}/(x\sqrt{s})<1$, $\hat k=(0,\sin\theta_0,\cos\theta_0)$, and 
$\theta_0$ is the polar angle of $\hat k$ with respect to the beam line, in the c.o.m. frame.
The subscript $_0$ denotes quantities evaluated in the c.o.m. frame.

The differential
 cross section    is generically given by
\begin{equation}
d\hat\sigma=\frac{\sum\overline{|{\cal M}|^2}}{4(p_1\cdot p_2)}
\de \Phi_3\,,
\end{equation}
where  the three-body phase space is
\be
\de \Phi_3 = (2\pi)^4 \, \delta(E_1+E_2-E_3-E_4-E_k) \, \delta^{(3)}(\vec p_1 +\vec p_2 - \vec p_3 - \vec p_4 - \vec k)
\frac{\de^3p_3}{(2\pi)^3 2E_3} \frac{\de^3p_4}{(2\pi)^3 2E_4} \frac{\de^3k}{(2\pi)^3 2E_k}.
\ee
Using the three-momentum delta function, we can integrate away $\de^3 p_4$; the energy delta function instead
fixes the angle $\theta_{0\,3{\rm j}}$ between $\hat p_3$ and the jet $\hat k$ as:
$ \cos\theta_{0\,3{\rm j}}=(\mathbf{p}_4^2-\mathbf{k}^2-\mathbf{p}_3^2)/{2|\mathbf{k}||\mathbf{p}_3|}$.
Integration  over the azimuthal angle $\phi_0$ of the outgoing jet simply gives a factor of $2\pi$, while the matrix element does depend on  the azimuthal angle of the three-momentum $\vec p_3$ with respect to $\vec k$, $\phi_{0\,3{\rm j}}$, and so it can not be integrated over at this stage,  contrary to the $s$-channel case.
Taking all of this into account, the phase space density simplifies to
\be\label{eq:phasespace}
\de \Phi_3 = \frac{1}{8(2\pi)^4} \, \de E_3 \, \de|\vec k| \, \de\cos\theta_0 \, \de\phi_{0\,3{\rm j}}
= \frac{x^2 s}{32(2\pi)^4} \, \de y_0 \, \de z_0 \, \de\cos\theta_0 \, \de\phi_{0\,3{\rm j}}.
\ee
The kinematical domains of $y_0$, $z_0$ and $\phi_{0\,3{\rm j}}$ are
\bea
\frac{z_0}{2}\left(1-\sqrt{\frac{1-z_0-a^2}{1-z_0}}\right) \leq & y_0 & \leq \frac{z_0}{2}\left(1+\sqrt{\frac{1-z_0-a^2}{1-z_0}}\right) \\
0\leq & z_0 & \leq 1-a^2 \\
0\leq & \phi_{0\,3{\rm j}} & \leq 2\pi
\eea
The variables $y_0$ and $\phi_{0\,3{\rm j}}$ refer to the momentum $\vec p_3$ of an invisible DM particle;
they are therefore not measurable, and we  integrate over them.
For our present purpose, finding the total integrated cross section is useless, since these variables enter our definition of the momentum transfer $Q_{\rm tr}$,
and the condition $Q_{\rm tr}<\Lambda$ which we used to define the ratio $R_\Lambda$.

With the matrix elements of Eqns. \ref{eq:Msqavgluon} and \ref{eq:Msqavquark}, and the phase space density \ref{eq:phasespace}, we get the differential cross sections in the c.o.m. frame:
\bea
\left.\frac{\de^4\hat\sigma}{\de z_0 \,\de\cos\theta_0\, \de y_0\,\de\phi_{0\,3{\rm j}}} \right|_g & = & 
\frac{1}{4608 \pi^4}\frac{g_s^2}{\Lambda^4}\frac{1-z_0}{z_0^4} \nn\\
&&\hspace{-3.5cm}\Biggl\{4 x (2-z_0) \csc\theta_0 \cos\phi_{0\,3{\rm j}} (\cos\theta_0 (z_0-2 y_0)+z_0) \sqrt{s \left(s x^2 y_0 (z_0-1) (y_0-z_0)-m_{\rm DM}^2 z_0^2\right)} \nn\\
&&\hspace{-3.5cm}-8 m_{\rm DM}^2 z_0^2 \cos ^2\phi_{0\,3{\rm j}}+
sx^2((z_0-2) z_0+2) \left(\sec ^2\left(\theta_0/2\right)y_0^2+\csc ^2\left(\theta_0/2\right) (y_0-z_0)^2\right) \nn\\
&&\hspace{-3.5cm}-2 sx^2 y_0^2  ((z_0-6) z_0+6)+4sx^2 y_0 (z_0-1) (y_0-z_0) \cos (2 \phi_{0\,3{\rm j}}) \nn\\
&&\hspace{-3.5cm}+2 sx^2 y_0 ((z_0-6) z_0+6) z_0 - sx^2 z_0^2 ((z_0-2) z_0+2)\Biggr\},
\eea
\bea
\left.\frac{\de^4\hat\sigma}{\de z_0 \,\de\cos\theta_0\, \de y_0\,\de\phi_{0\,3{\rm j}}} \right|_q & = &
   \frac{1}{98304\pi^4}\frac{g_s^2}{\Lambda^4}\frac{1-z_0}{z_0^3 \cos^2\frac{\theta_0}{2}}  \nn\\
&&\hspace{-3.5cm}\Biggl\{
   8 x \sqrt{s}\left[ z_0 (z_0 - y_0 -1) - \left(z_0^2-(1+y_0)z_0+2y_0 \right) \cos\theta_0 \right] \cos\phi_{0\,3{\rm j}} \sin\theta_0 \times \nn\\
&&\hspace{-2.5cm}
          \times \sqrt{ s x^2 y_0 (z_0 - y_0) (1 - z_0) - m^2_{\rm DM} z_0^2} \nn\\
&&\hspace{-3.5cm}
   - 2 (1-\cos(2 \theta_0)) m^2_{\rm DM} z_0^2
   + 4 \Bigl[s x^2 y_0 (z_0 - y_0) (1 - z_0) - m^2_{\rm DM} z_0^2\Bigr] \cos(2\phi_{0\,3{\rm j}})\sin^2\theta_0 \nn\\
&&\hspace{-3.5cm}
   + s x^2 \Bigl[ 11z_0^4 - (6+22y_0)z_0^3 + (11y_0^2+8y_0+3)z_0^2 - 2y_0(1+y_0)z_0 + 2y_0^2 \Bigr] \nn\\
&&\hspace{-3.5cm}
   + s x^2 \Bigl[ z_0^4 - 2(1+y_0)z_0^3 + (y_0^2+8y_0+1)z_0^2 - 6y_0(1+y_0)z_0 + 6y_0^2 \Bigr] \cos(2 \theta_0) \nn\\
&&\hspace{-3.5cm}
   - 4 s x^2 z_0 \Bigl[ z_0^3 - 2(1+y_0)z_0^2 + (y_0^2+4y_0+1)z_0 -2y_0(1+y_0) \Bigr] \cos\theta_0
\Biggr\} . \nn\\
\eea

To get the cross sections in the lab frame we perform a boost in the $\hat z$ axis, accounting for the generic parton momentum fractions $x_1$, $x_2$.
The velocity of the c.o.m. of the colliding particles with respect to the lab frame is given by
\be
\beta_{\rm c.o.m.} = \frac{x_1-x_2}{x_1+x_2},
\ee
so that the relations between the quantities $z_0$, $\theta_0$ and the analogous ones $z$, $\theta$ in the lab frame are
\bea\label{eq:boost}
z_0 & = & \frac{(x_1+x_2)^2+(x_2^2-x_1^2)\cos\theta}{4x_1x_2} z \nn \\
\sin^2\theta_0 & = & \frac{4x_1x_2}{[(x_1+x_2)+(x_2-x_1)\cos\theta]^2} \sin^2\theta.
\eea
The Jacobian factor to transform $\de z_0\,\de\cos\theta_0 \to \de z\,\de\cos\theta$ is simply obtained using equations \ref{eq:boost};
the cross section in the lab frame is then
\be
\frac{\de^4\hat\sigma}{\de z \,\de\cos\theta\, \de y_0\,\de\phi_{0\,3{\rm j}}} =
\frac{x_1+x_2}{x_1+x_2+(x_1-x_2)\cos\theta}
\left. \frac{\de^4\hat\sigma}{\de z_0 \,\de\cos\theta_0\, \de y_0\,\de\phi_{0\,3{\rm j}}} \right\vert_{
\begin{footnotesize}
\begin{array}{l}
z_0\to z_0(z)\\
\theta_0\to \theta_0(\theta)
\end{array}
\end{footnotesize}
}\,.
\ee
Expressing the energy of the emitted gluon or (anti-)quark in terms of the transverse momentum and rapidity,
$k^0 = p_{\rm T}\cosh\eta$, one finds
\be
z=\frac{4p_{\rm T}\cosh\eta}{(x_1+x_2)\sqrt{s}}, \qquad \cos\theta=\tanh\eta
\ee
which allows us to express the differential cross sections with respect to the transverse momentum and pseudo-rapidity of the emitted jet:
\be
\frac{\de^4\hat\sigma}{\de p_{\rm T} \,\de\eta\, \de y_0\,\de\phi_{0\,3{\rm j}}} =
\frac{4}{(x_1+x_2)\sqrt{s}\cosh\eta}
\left. \frac{\de^4\hat\sigma}{\de z \,\de\cos\theta\, \de y_0\,\de\phi_{0\,3{\rm j}}} \right\vert_{
\begin{footnotesize}
\begin{array}{l}
z\to z(p_{\rm T},\eta)\\
\theta\to \theta(p_{\rm T},\eta)
\end{array}
\end{footnotesize}
}\,.
\ee

\subsection{Transferred momentum\label{app:Qtr}}
As is clear from  our arguments, the key ingredient to quantify the validity of the EFT approximation is the value of the transferred momentum of the process.
Since each process of interest here is given (at tree level) by the contribution of two Feynman diagrams, there will also be two expressions for the transferred momentum
for both gluon and (anti-)quark emission, which we report here:
\bea
Q_{\rm tr,g1}^2 & = & (p_1-k-p_3)^2 \nn\\
&=&m_{\rm DM}^2 + \sqrt{s} x_2 e^\eta p_{\rm T} - \frac{e^{2\eta}(1+y)(x_1x_2^2s)}{x_1+e^{2\eta}x_2} - \frac{x_1^2x_2^2e^\eta s^{3/2} y\left(x_1-e^{2\eta}x_2\right)}{p_{\rm T}(x_1+e^{2\eta}x_2)^2} \nn \\
&&-\frac{2 e^{\eta } x_1 x_2 \sqrt{s} \cos\phi_{\rm 0\,3{\rm j}}}{p_{\rm T}(x_1+e^{2\eta}x_2)^2}\Bigl[ -m_{\rm DM}^2 p_{\rm T}^2 \left(x_1+e^{2 \eta } x_2\right)^2 \\
&& {- s x_1 x_2 y \left(e^{\eta} \sqrt{s} x_1 x_2-p_{\rm T} \left(x_1+e^{2 \eta } x_2\right)\right) \left(e^{\eta } \sqrt{s} x_1 x_2 y-p_{\rm T} \left(x_1+e^{2 \eta } x_2\right)\right)\Bigr]}^{1/2} \Biggr\}, \nn
\eea
\bea
Q_{\rm tr,g2}^2 & = & (p_1-p_3)^2 \nn\\
&=&m_{\rm DM}^2
+ \frac{x_1x_2s(x_1-e^{2\eta}x_2)}{x_1+e^{2\eta}x_2}
- \frac{(1-y)(x_1^2x_2s)}{x_1+e^{2\eta}x_2}
- \frac{x_1^2x_2^2 e^{\eta}s^{3/2}y (x_1-e^{2\eta} x_2)}{p_{\rm T}(x_1+e^{2\eta} x_2)^2} \nn\\
&&-\frac{2 e^{\eta } x_1 x_2 \sqrt{s} \cos\phi_{\rm 0\,3{\rm j}}}{p_{\rm T}(x_1+e^{2\eta}x_2)^2}\Bigl[ -m_{\rm DM}^2 p_{\rm T}^2 \left(x_1+e^{2 \eta } x_2\right)^2 \\
&& {- s x_1 x_2 y \left(e^{\eta} \sqrt{s} x_1 x_2-p_{\rm T} \left(x_1+e^{2 \eta } x_2\right)\right) \left(e^{\eta } \sqrt{s} x_1 x_2 y-p_{\rm T} \left(x_1+e^{2 \eta } x_2\right)\right)\Bigr]}^{1/2} \Biggr\}, \nn
\eea
\bea
Q_{\rm tr,q1}^2 & = & (p_3+k)^2 \nn\\
&=& m_{\rm DM}^2 + p_{\rm T}\sqrt{s}\left(e^{-\eta}x_1 + e^{\eta}x_2\right) - x_1x_2s\,y ,
\eea
\bea
Q_{\rm tr,q2}^2 & = & (p_1-p_3-k)^2 \nn\\
&=& m_{\rm DM}^2 
+ \sqrt{s} x_1  e^{-\eta}p_{\rm T} 
- \frac{(1+y) (x_1^2 x_2 s)}{x_1+e^{2\eta}x_2} 
+ \frac{x_1^2x_2^2 e^{\eta}s^{3/2}y (x_1-e^{2\eta} x_2)}{p_{\rm T}(x_1+e^{2\eta} x_2)^2} \nn\\
&&-\frac{ 2 e^{\eta } x_1 x_2 \sqrt{s} \cos\phi_{\rm 0\,3{\rm j}} }{p_{\rm T}(x_1+e^{2\eta}x_2)^2}
\Bigl[-m_{\rm DM}^2 p_{\rm T}^2 \left(x_1+e^{2 \eta } x_2\right)^2 \\
&&{- s x_1 x_2 y \left(e^{\eta } \sqrt{s} x_1 x_2-p_{\rm T} \left(x_1+e^{2 \eta } x_2\right)\right) \left(e^{\eta } \sqrt{s} x_1 x_2 y-p_{\rm T} \left(x_1+e^{2 \eta } x_2\right)\right)\Bigr]}^{1/2}.\nn
\eea
The notation $g,q$ stands for gluon or quark emission; the indices $1,2$ refer to emission from each of the initial state particles.

\bibliographystyle{h-physrev5}
\bibliography{bibfile}

\begin{thebibliography}{10}

\bibitem{Baudis:2012ig}
L.~Baudis,
\newblock Phys.Dark Univ. {\bf 1}, 94 (2012), arXiv:1211.7222.

\bibitem{Cirelli:2012tf}
M.~Cirelli,
\newblock Pramana {\bf 79}, 1021 (2012), arXiv:1202.1454.

\bibitem{Feng:2014uja}
J.~Feng {\em et~al.},
\newblock (2014), arXiv:1401.6085.

\bibitem{Abramowski:2011hc}
H.E.S.S.Collaboration, A.~Abramowski {\em et~al.},
\newblock Phys.Rev.Lett. {\bf 106}, 161301 (2011), arXiv:1103.3266.

\bibitem{Ackermann:2013yva}
Fermi-LAT Collaboration, M.~Ackermann {\em et~al.},
\newblock Phys.Rev. {\bf D89}, 042001 (2014), arXiv:1310.0828.

\bibitem{Aprile:2012nq}
XENON100 Collaboration, E.~Aprile {\em et~al.},
\newblock Phys.Rev.Lett. {\bf 109}, 181301 (2012), arXiv:1207.5988.

\bibitem{Akerib:2013tjd}
LUX Collaboration, D.~Akerib {\em et~al.},
\newblock (2013), arXiv:1310.8214.

\bibitem{Bernabei:2010mq}
DAMA Collaboration, LIBRA Collaboration, R.~Bernabei {\em et~al.},
\newblock Eur.Phys.J. {\bf C67}, 39 (2010), arXiv:1002.1028.

\bibitem{Agnese:2013rvf}
CDMS Collaboration, R.~Agnese {\em et~al.},
\newblock Phys.Rev.Lett. {\bf 111}, 251301 (2013), arXiv:1304.4279.

\bibitem{Aalseth:2012if}
CoGeNT Collaboration, C.~Aalseth {\em et~al.},
\newblock Phys.Rev. {\bf D88}, 012002 (2013), arXiv:1208.5737.

\bibitem{monojetATLAS1}
ATLAS Collaboration, G.~Aad {\em et~al.},
\newblock JHEP {\bf 1304}, 075 (2013), arXiv:1210.4491.

\bibitem{monojetCMS1}
CMS Collaboration, S.~Chatrchyan {\em et~al.},
\newblock Phys.Rev.Lett. {\bf 107}, 201804 (2011), arXiv:1106.4775.

\bibitem{monojetATLAS2}
ATLAS Collaboration,
\newblock (2012).

\bibitem{monojetCMS2}
CMS Collaboration,
\newblock (2013).

\bibitem{ATLASWZ}
T.~A. collaboration,
\newblock (2013).

\bibitem{Aad:2014vka}
ATLAS Collaboration, G.~Aad {\em et~al.},
\newblock (2014), arXiv:1404.0051.

\bibitem{monogammaATLAS1}
ATLAS Collaboration, G.~Aad {\em et~al.},
\newblock Phys.Rev.Lett. {\bf 110}, 011802 (2013), arXiv:1209.4625.

\bibitem{monogammaCMS1}
CMS Collaboration, S.~Chatrchyan {\em et~al.},
\newblock Phys.Rev.Lett. {\bf 108}, 261803 (2012), arXiv:1204.0821.

\bibitem{monogammaATLAS2}
ATLAS Collaboration,
\newblock (2012).

\bibitem{monogammaCMS2}
CMS Collaboration, C.~Collaboration,
\newblock (2011).

\bibitem{deSimone:2014pda}
A.~De~Simone, G.~F. Giudice, and A.~Strumia,
\newblock (2014), arXiv:1402.6287.

\bibitem{Alves:2013tqa}
A.~Alves, S.~Profumo, and F.~S. Queiroz,
\newblock JHEP {\bf 1404}, 063 (2014), arXiv:1312.5281.

\bibitem{Busoni:2013lha}
G.~Busoni, A.~De~Simone, E.~Morgante, and A.~Riotto,
\newblock Phys.Lett. {\bf B728}, 412 (2014), arXiv:1307.2253.

\bibitem{Busoni:2014sya}
G.~Busoni, A.~De~Simone, J.~Gramling, E.~Morgante, and A.~Riotto,
\newblock (2014), arXiv:1402.1275.

\bibitem{Fox:2011pm}
P.~J. Fox, R.~Harnik, J.~Kopp, and Y.~Tsai,
\newblock Phys.Rev. {\bf D85}, 056011 (2012), arXiv:1109.4398.

\bibitem{Buchmueller:2013dya}
O.~Buchmueller, M.~J. Dolan, and C.~McCabe,
\newblock JHEP {\bf 1401}, 025 (2014), arXiv:1308.6799.

\bibitem{Bell:2012rg}
N.~F. Bell {\em et~al.},
\newblock Phys.Rev. {\bf D86}, 096011 (2012), arXiv:1209.0231.

\bibitem{Chang:2013oia}
S.~Chang, R.~Edezhath, J.~Hutchinson, and M.~Luty,
\newblock Phys.Rev. {\bf D89}, 015011 (2014), arXiv:1307.8120.

\bibitem{An:2013xka}
H.~An, L.-T. Wang, and H.~Zhang,
\newblock (2013), arXiv:1308.0592.

\bibitem{Bai:2013iqa}
Y.~Bai and J.~Berger,
\newblock JHEP {\bf 1311}, 171 (2013), arXiv:1308.0612.

\bibitem{DiFranzo:2013vra}
A.~DiFranzo, K.~I. Nagao, A.~Rajaraman, and T.~M.~P. Tait,
\newblock JHEP {\bf 1311}, 014 (2013), arXiv:1308.2679.

\bibitem{Papucci:2014iwa}
M.~Papucci, A.~Vichi, and K.~M. Zurek,
\newblock (2014), arXiv:1402.2285.

\bibitem{Garny:2014waa}
M.~Garny, A.~Ibarra, S.~Rydbeck, and S.~Vogl,
\newblock (2014), arXiv:1403.4634.

\bibitem{Bell:2010ei}
N.~F. Bell, J.~B. Dent, T.~D. Jacques, and T.~J. Weiler,
\newblock Phys.Rev. {\bf D83}, 013001 (2011), arXiv:1009.2584.

\bibitem{Goodman:2010ku}
J.~Goodman {\em et~al.},
\newblock Phys.Rev. {\bf D82}, 116010 (2010), arXiv:1008.1783.

\bibitem{pdf1}
A.~Martin, W.~Stirling, R.~Thorne, and G.~Watt,
\newblock Eur.Phys.J. {\bf C63}, 189 (2009), arXiv:0901.0002.

\bibitem{pdf3}
http://mstwpdf.hepforge.org/ .

\bibitem{Bertone:2004pz}
G.~Bertone, D.~Hooper, and J.~Silk,
\newblock Phys.Rept. {\bf 405}, 279 (2005), arXiv:hep-ph/0404175.

\bibitem{Ade:2013zuv}
Planck Collaboration, P.~Ade {\em et~al.},
\newblock (2013), arXiv:1303.5076.

\bibitem{Christensen:2008py}
N.~D. Christensen and C.~Duhr,
\newblock Comput.Phys.Commun. {\bf 180}, 1614 (2009), arXiv:0806.4194.

\bibitem{mg5}
J.~Alwall, M.~Herquet, F.~Maltoni, O.~Mattelaer, and T.~Stelzer,
\newblock JHEP {\bf 1106}, 128 (2011), arXiv:1106.0522.

\bibitem{Pumplin:2002vw}
J.~Pumplin {\em et~al.},
\newblock JHEP {\bf 0207}, 012 (2002), arXiv:hep-ph/0201195.

\end{thebibliography}

\end{document}